\documentclass[
	aps, prd, reprint,
	10pt, notitlepage, a4paper,
        floats, floatfix,
	amsmath, amssymb, amsfonts, eqsecnum,
	superscriptaddress,
	showpacs, showkeys,
	nofootinbib,
 	longbibliography,
]{revtex4-1}

\usepackage{aas_macros}
\usepackage[usenames,dvipsnames]{xcolor}
\usepackage{xspace} % Sensible space treatment at end of simple macros
\usepackage{bm,graphicx} % bold math
\usepackage[utf8]{inputenc} % for some references from inspirehep.net

\xdefinecolor{mylinkcolor}{rgb}{0,0,0.5}
\usepackage[
	bookmarksnumbered, bookmarksopen, bookmarksopenlevel=2,
	breaklinks=true,
	colorlinks=true, filecolor=mylinkcolor, citecolor=mylinkcolor,
	linkcolor=mylinkcolor, urlcolor=mylinkcolor, menucolor=mylinkcolor,
]{hyperref}

\def\vct#1{{\bm{#1}}}

\def\nl{\\ & \quad}
\def\nnl{\nonumber \\ & \quad}

\def\ee{\end{equation}}
\def\eea{\end{eqnarray}}
\def\be{\begin{equation}}
\def\bea{\begin{eqnarray}}

\DeclareMathOperator{\Order}{\mathcal{O}}

\def\Lcov{\Lambda}

\allowdisplaybreaks

\begin{document}

\newcommand{\AEI}{\affiliation{Max-Planck-Institute for Gravitational Physics (Albert-Einstein-Institute),
\\ Am M{\"u}hlenberg 1, 14476 Potsdam-Golm, Germany, EU}}

\newcommand{\GRAPPA}{\affiliation{GRAPPA Institute of High-Energy Physics, University of Amsterdam, Science Park 904, 1098 XH Amsterdam, The Netherlands, EU}}

\newcommand{\DITF}{\affiliation{Delta Institute for Theoretical Physics, Science Park 904, 1090 GL Amsterdam, The Netherlands, EU}}

\newcommand{\ITF}{\affiliation{Institute for Theoretical Physics, Utrecht University, Princetonplein 5, 3584 CC Utrecht, The Netherlands, EU}}

\newcommand{\UU}{\affiliation{Department of Physics, Utrecht University, Princetonplein 1, 3584 CC Utrecht, The Netherlands, EU}}

\newcommand{\GRASP}{\affiliation{Institute for Gravitational and Subatomic Physics (GRASP),\\ Department of Physics, Utrecht University, Princetonplein 1, 3584 CC Utrecht, The Netherlands, EU}}

\newcommand{\Nikhef}{\affiliation{Nikhef, Science Park, 1098 XG Amsterdam, The Netherlands, EU}}

\title{Relativistic effective action of dynamical gravitomagnetic tides\\for slowly rotating neutron stars}

\date{\today }

\author{Pawan Kumar Gupta}
\email{p.gupta@nikhef.nl}
%\homepage{}
\Nikhef \GRASP

\author{Jan Steinhoff}
\email{jan.steinhoff@aei.mpg.de}
\homepage{http://jan-steinhoff.de/physics/}
\AEI

\author{Tanja Hinderer}
\email{t.p.hinderer@uu.nl}
%\homepage{}
\ITF \GRAPPA \DITF

\begin{abstract}
 Gravitomagnetic quasi-normal modes of neutron stars are resonantly excited by tidal effects during a binary inspiral, leading to a potentially measurable effect in the gravitational-wave signal. We take an important step towards incorporating these effects in waveform models by developing a relativistic effective action for the gravitomagnetic dynamics that clarifies a number of subtleties. Working in the slow-rotation limit, we first consider the post-Newtonian approximation and explicitly derive the effective action from the equations of motion. We demonstrate that this formulation opens a way to compute mode frequencies, yields insights into the relevant matter variables, and elucidates the role of a shift symmetry of the fluid properties under a displacement of the gravitomagnetic mode amplitudes. We then construct a fully relativistic action based on the symmetries and a power counting scheme. This action involves four coupling coefficients that depend on the internal structure of the neutron star and characterize the key matter parameters imprinted in the gravitational waves. We show that, after fixing one of the coefficients by normalization, the other three directly involve the two kinds of gravitomagnetic Love numbers (static and irrotational), and the mode frequencies. We discuss several interesting features and dynamical consequences of this action, and analyze the frequency-domain response function (the frequency-dependent ratio between the induced flux quadrupole and the external gravitomagnetic field), and a corresponding Love operator representing the time-domain response. Our results provide the foundation for deriving precision predictions of gravitomagnetic effects, and the nuclear physics they encode, for gravitational-wave astronomy. 
\end{abstract}

\maketitle

%\tableofcontents

%%%%%%%%%%%%%%%%%%%%%%%%%%%%%%%%%%%%%%%
\section{Introduction}
%%%%%%%%%%%%%%%%%%%%%%%%%%%%%%%%%%%%%%%

%general statements on GWs from inspirals, probing NS matter, dynamical tides, relation to tidal deformability
Gravitational waves from inspiraling binary neutron stars encode unique information on the matter at supra-nuclear densities in their interiors \cite{Abbott:2018exr, Abbott:2018wiz, Weinberg:2018icl, TheLIGOScientific:2017qsa, Abbott:2020uma, Abbott:2020khf}. Understanding the properties of matter at such extreme density remains an important frontier in subatomic physics~\cite{Nupecc,Geesaman:2015fha}. Among the most interesting imprints of matter on the gravitational waves during a binary inspiral are signatures of tidal effects. Tidal effects comprise a rich set of phenomena associated with the excitation of the stars' quasi-normal modes. The mode excitation can be either resonant or adiabatic, depending on the rate of variations in the tidal fields due to the spacetime curvature produced by the orbiting companion compared to the characteristic mode frequency. The excitation of quasi-normal modes is most commonly considered for the ringdown signals in black hole binaries, where the merger excites a broad spectrum of quasi-normal modes of the remnant which  damp away due to gravitational radiation. Neutron stars have a much richer mode spectrum than black holes due to the presence of matter. Several classes of neutron star modes have sufficiently low frequencies to become individually excited during a binary inspiral. This opens the possibility for a detailed, spectroscopic characterization of the ground-state matter in neutron-star interiors from gravitational waves emitted during their inspiral, provided that the modes have sufficiently large tidal coupling strengths to lead to a noticeable effect. The fundamental modes typically have the largest tidal couplings. They are an example of gravitoelectric phenomena associated with the tidal deformability, a parameter that is measurable in the gravitational-wave signals~\cite{Abbott:2017dke,Abbott:2018wiz,TheLIGOScientific:2017qsa}. There are also several other interesting classes of modes predominantly connected with gravitoelectric tides~\cite{Kokkotas:1999bd}. 

%gravitomagnetic tides and magnetic sector of inertial modes
An intriguing feature of general relativity is the emergence of new types of gravitomagnetic tides, which have no Newtonian analogs. Gravitomagnetic tides most strongly excite the magnetic (odd-parity) sector of inertial modes of a rotating star. Inertial modes are associated with the Coriolis effect and include the $r$-modes~\cite{1981A&A....94..126P,Ho:1998hq,Schenk:2001zm,Lockitch:1998nq}, which are inertial modes with purely magnetic parity. The $r$-modes have received significant attention due to their unusual properties and the fact that they can become unstable to gravitational radiation (see, e.g., Refs.~\cite{Friedman:1997uh, Andersson:1997xt,Kokkotas:2015gea,Flanagan:2006sb,Schenk:2001zm,Poisson:2020eki,Poisson:2020mdi,Poisson:2020ify,Ma:2020oni}). The remaining inertial modes have no specific name and are of mixed parity. In the slow-rotation limit, only their magnetic parts are directly relevant for gravitomagnetic tides. The inertial mode frequencies are approximately proportional to the rotation frequency of the star. Consequently, the gravitomagnetic inertial-mode resonances in a binary generally lie well within the sensitive frequency band of gound-based gravitational-wave detectors \cite{TheLIGOScientific:2014jea, TheVirgo:2014hva, Aso:2013eba, LIGOIndia}. This opens interesting prospects for probing properties of neutron-star matter beyond the information encoded in gravitoelectric tidal deformability or radius. However, it is currently not possible to measure this new physics because a relativistic modeling framework of gravitomagnetic dynamical tides is not developed. 

%effect of modes on the dynamics, key parameters in GWs, gravitomagnetic Love
The tidal excitation of a quasi-normal mode in quasi-circular binaries is analogous to a harmonic oscillator with a quasi-periodic force. The effect of quadrupolar gravitomagnetic tidal mode excitation was estimated in Refs.~\cite{Flanagan:2006sb, Poisson:2020eki, Poisson:2020mdi,Ma:2020oni}; see Refs.~\cite{Xu:2017hqo,Ho:1998hq,Lai:2006pr} for studies based on the weaker coupling to gravitoelectric fields. These results indicate that the impact on the gravitational-wave phasing is large enough to be potentially measurable with the planned future upgrades to current detectors and third-generation facilities. The gravitational-wave signatures from mode excitations directly depend on key matter parameters: the Love numbers characterizing how strongly the mode couples to the tidal field, and the mode frequency. These parameters are computed from linearized perturbations to a relativistic star in equilibrium. The gravitomagnetic mode frequencies were obtained, e.g., in Refs.~\cite{Kokkotas:1999bd, Andersson:2000mf,Idrisy:2014qca,Lee:2002fp,Kokkotas:2015gea}. The Love numbers, however, require taking the limit that the perturbing frequency goes to zero, which has proved subtle, and leads to two distinct Love numbers. They are associated with the different assumptions of a static or irrotational perturbed fluid. These unusual features of the response of a neutron star to a gravitomagnetic tidal perturbation have prompted several discussions in the literature~\cite{Damour:2009vw, Binnington:2009bb, Landry:2015cva, Landry:2015snx, Poisson:2016wtv, Pani:2018inf}, and were recently re-examined in the context of a post-Newtonian star in Refs.~\cite{Poisson:2020eki, Poisson:2020mdi, Poisson:2020ify}.

%This paper and its impact
 The promising prospects for measuring the gravitomagnetic modes motivate the need for modern gravitational-wave models to include these phenomena. A crucial foundation for developing state-of-the-art waveform models of matter effects in binary inspirals is a relativistic effective action for the dynamics. The Love numbers and mode frequencies immediately appear in the coupling coefficients in this effective action. In this paper, we derive a relativistic effective action for gravitomagnetic tidal effects in the slow-rotation limit. We develop the theory by first considering a post-Newtonian approximation of the neutron-star interior. This enables us to identify a new way to compute the mode frequencies from the perspective of a rotation-induced shift away from its vanishing value for nonrotating stars. It also yields important insights into the relevant matter variables for the dynamics, and their connection to the mode functions. Further, these studies reveal the important role of a shift symmetry, whereby a displacement of the gravitomagnetic mode amplitudes leaves the global properties of the fluid unchanged. Ensuring that the action respect this symmetry has direct consequences for its formulations in the corotating and inertial frames. 
 
 Next, we develop the fully relativistic theory based on the symmetries and a power counting scheme. We find that within our approximations, the dominant effects are described by four nontrivial couplings that come with coefficients that encode the microphysics of neutron-star interiors. We discuss the matching of these coefficients to the relativistic magnetic tidal deformabilities (Love numbers) and mode frequencies. Notably, we show that both kinds of magnetic Love numbers, the static and irrotational ones, appear in the action and are thus relevant for gravitational waves. The static Love number corresponds to the coefficient of a nonlinear field contribution, as discussed in the post-Newtonian context in Ref.~\cite{Poisson:2020ify}. The difference between static and irrotational Love numbers characterizes the direct contribution from the magnetic modes. To identify and match the mode frequency we calculate the relativistic response function and discuss its features.  We also derive its limiting form in several regimes after clarifying various subtleties, and obtain a corresponding Love operator representing the time-domain response. Our action provides a key foundation for accurately modeling  gravitomagnetic effects in gravitational waves and interpreting the information on subatomic physics they encode. 

%organization with some explanation 
The paper is organized as follows.
In Sec.~\ref{sec:Newtaction} we review the treatment of dynamical gravitomagnetic tides in a first post-Newtonian (1PN) approximation from Ref.~\cite{Flanagan:2006sb}. We work to linear order in the rotation of the star and derive an action that encodes the excitation of magnetic modes by an external gravitomagnetic field. We discuss the relevant degrees of freedom and their relation to contributions from individual modes. We also highlight the shift symmetry that occurs in the gravitomagnetic sector and its importance for the Lagrangians in the corotating and inertial frames. 
In Sec.~\ref{sec:relativistic} we construct a fully relativistic action in the framework of effective field theory. We briefly discuss the power counting scheme, and further specialize to the four interaction terms that are most important based on the post-Newtonian limit. In
Sec.~\ref{sec:matching} we perform the matching of the coefficients in the effective action to the relativistic mode frequencies and tidal deformabilities.
Section~\ref{sec:response} discusses the  frequency- and time-domain tidal response and their limiting forms in different regimes.
In Sec.~\ref{sec:summary} we provide a brief summary of the relativistic Lagrangian and the coupling coefficients involved.
We also discuss the physical insights and dynamical consequences of this action.
Section~\ref{sec:conclusions} summarizes our conclusions, and Appendix~\ref{sec:appendix} contains a short compilation of useful formulas.

%notation and conventions
We use geometric units with $G=c=1$, with $G$ being the gravitational constant and $c$ the speed of light, except in cases where we make the post-Newtonian counting explicit as a formal expansion in $c^{-2}$. We denote spatial tensors expressed in the corotating frame of the star by capital Latin letters $I,J,K,\ldots$, and use boldface notation for three-dimensional vectors in this frame. We use lowercase letters $i,j,k,\ldots$ for the inertial frame. Greek letters denote four-dimensional spacetime coordinate indices in the inertial frame, and a calligraphic index ${\cal B}$  indicates the magnetic part of a quantity.
Our convention for the Riemann tensor is
\begin{equation}
R^{\mu}{}_{\nu\alpha\beta} = \Gamma^{\mu}{}_{\nu \beta , \alpha}
        - \Gamma^{\mu}{}_{\nu \alpha , \beta}
        + \Gamma^{\rho}{}_{\nu \beta} \Gamma^{\mu}{}_{\rho \alpha}
        - \Gamma^{\rho}{}_{\nu \alpha} \Gamma^{\mu}{}_{\rho \beta} ,
\end{equation}
where $\Gamma^{\mu}{}_{\nu \beta}$ is the Christoffel symbol and the comma denotes a partial derivative.

%%%%%%%%%%%%%%%%%%%%%%%%%%%%%%%%%%%%%%%
\section{Dynamical magnetic tides of rotating stars}
\label{sec:Newtaction}
%%%%%%%%%%%%%%%%%%%%%%%%%%%%%%%%%%%%%%%

In this section, we briefly review the description of gravitomagnetic modes of a neutron star in the presence of an external gravitomagnetic field from Ref.~\cite{Flanagan:2006sb}. We work within a double perturbative expansion in the post-Newtonian and slow-rotation approximations. From the equations of motion we develop an effective action in the corotating frame of the neutron star. The Lagrangian formulation provides a clean, elegantly concise description of the unusual features of the gravitomagnetic dynamics compared to the more familiar gravitoelectric tides. We formulate the action in terms of symmetric-tracefree tensors of magnetic parity, which conveniently isolates the relevant contributions from associated modes into effective degrees of freedom and elucidates the underlying physics of gravitomagnetic tides. We give an explicit example demonstrating the utility of the Lagrangian by exhibiting a way to calculate the gravitomagnetic mode frequency. We also discuss the important role of a shift symmetry, whereby a displacement in the mode amplitudes leaves the fluid properties unchanged. Requiring that the Lagrangian respect this symmetry has an impact on its form in the inertial frame. The insights on the matter variables and symmetries developed in this section will be important for constructing the relativistic theory in Sec.~\ref{sec:relativistic}.

\subsection{Metric of a slowly rotating neutron star}

We consider a (approximately) spherical neutron star of mass $M$ that is slowly rotating with angular velocity $\Omega$ and immersed in an external gravitomagnetic tidal potential of 1PN order. The post-Newtonian approximation can be understood as a formal expansion in the squared inverse of the speed of light $c$. To highlight similarities with electromagnetism, it is convenient to write the metric for the neutron star in the inertial\footnote{We understand here an inertial frame in the global Newtonian sense. The metric asymptotically approaches the Minkowski one in the inertial frame.} frame, denoted by indices $i$, $j$, $k$, \dots, in the form~\cite{Kol:2007bc, Damour:1990pi, Kaplan:2008dh}
\begin{equation}
\label{eq:metricinertial}
\begin{aligned} d s^{2}=&-\exp \left[ \frac{2\phi}{c^2} \right] \left[ c\,  d t - \frac{1}{c^3}  A_i d x^i \right]^2  \\ & + \exp \left[ - \frac{2 \phi}{c^2} \right] \gamma_{ij}\,  d x^i d x^j ,
\end{aligned}
\end{equation}
where $\phi$ is the gravitoelectric (Newtonian) potential, $A^i$ is the gravitomagnetic potential, and to 1PN order  $\gamma_{ij} = \delta_{ij} + \Order (c^{-4})$. We next make a spatial coordinate change that keeps $t$ unchanged from the inertial frame $x^i$ to the corotating frame $x^I$ (denoted by capitalized indices $I$, $J$, $K$, \dots). This transformation is given by
\begin{equation}
\label{eq:coordchange}
x^j = R_I{}^j x^I , \quad R_I{}^j = C_J{}^j \exp (*{\bm\Omega} \, t)_I{}^J ,
\end{equation}
where $\bm \Omega$ is the angular velocity vector, $*\bm \Omega_{IJ} \equiv \Omega_{IJ}= \epsilon_{IJK} \Omega^K$ is its antisymmetric dual tensor, $\epsilon_{IJK}$ the Levi-Civita symbol, $R_I{}^j$ is a rotation matrix ($R_I{}^k R_{Jk} = \delta_{IJ}$) expressed here using a matrix exponential of $*{\bm\Omega} \, t$, and $C_I{}^j$ a constant rotation matrix identical to $R_I{}^j$ at $t=0$.
We adopt the convention that boldface notation for spatial vectors refers to components in the corotating frame, e.g., $\bm \Omega = (\Omega^I)$, and that spatial indices are raised and lowered using the Kronecker delta.
The angular velocity can be expressed as
\begin{equation}
  \label{eq:OmegaNewton}
  \Omega^I = \frac{1}{2} \epsilon^{IJK} \Omega_{JK}, \quad \Omega_{JK} = - \Omega_{KJ} = \dot{R}_J{}^i R_{Ki} .
\end{equation}
Applying this coordinate change to the line element \eqref{eq:metricinertial} is straightforward. The differentials of the coordinates are related by
\begin{equation}
  d x^j = R_I{}^j [ d x^I - (\bm x \times \bm \Omega)^I dt ] ,
\end{equation}
which leads to the corotating-frame 1PN line element
\begin{equation}
\label{eq:metriccorot}
\begin{aligned}
  d s^{2}=&-\left[c^2+ 2\phi + \frac{2\phi^2}{c^2} -\frac{2}{c^2}{\bm\Omega}\cdot({\bm x}\times {\bm A})\right] d t^{2} \\
  &+2\left[ - \bm x \times \bm \Omega \left( 1-\frac{2 \phi}{c^2} \right) + \frac{\bm A}{c^2} \right] \cdot d \bm x \, d t \\
  &+ \left[1-\frac{2 \phi}{c^2} \right] d \bm x \cdot  d \bm x + \Order(c^{-4}, \Omega^2).
\end{aligned}
\end{equation}
We have only kept terms linear in the angular velocity, since we are interested in slowly rotating stars.
In the next section, we use the metric in Eq.~\eqref{eq:metriccorot} to obtain the Euler equation for the matter inside the slowly rotating neutron star.

\subsection{Fluid perturbation in the corotating frame}
We describe the matter inside the neutron star as a perfect fluid with energy-momentum tensor 
\begin{equation}
\label{eq:Tmunuprime}
T^{\mu\nu} = \left( \rho + \frac{p}{c^2} \right) u^{\mu} u^{\nu} + p \, g^{\mu\nu} ,
\end{equation}
where $\rho$ is the mass density, $p$ is the pressure, and $u^\mu$ is the 4-velocity of the fluid normalized as $u_\mu u^\mu = - c^2$. The neutron-star matter is subject to energy-momentum conservation $T^{\mu\nu}{}_{;\nu} = 0$, where the semicolon denotes the covariant derivative. Evaluating the energy-momentum conservation using the metric in Eq.~\eqref{eq:metriccorot} leads to the corotating-frame Euler equation in Lagrangian form given by
\begin{equation}
\label{eq:Eulereq}
\dot{\bm u}+2 \bm\Omega \times {\bm u}+ {\bm u} \cdot \bm \nabla {\bm u}=-\frac{\bm \nabla p}{\rho}-\bm \nabla \phi+ \frac{\bm \zeta}{c^2} + \dots .
\end{equation}
 Here, at 1PN order ($c^{-2}$) we show only the terms involving the gravitomagnetic potential defined by
\begin{equation}
\label{eq:zetadef}
{\bm \zeta}=-\dot{\bm A}-{\bm\Omega} \times {\bm A}+({\bm \Omega} \times {\bm x}) \cdot \bm \nabla {\bm A}+({\bm u}+{\bm \Omega} \times {\bm x}) \times({\bm \nabla} \times {\bm A}),
\end{equation}
where the overdot denotes a time derivative $\dot{~} = \partial / \partial t$.
Recall that we also work to linear order in the angular velocity. This is the reason for the absence of the centripetal force, which is quadratic in $\bm \Omega$, from Eq.~\eqref{eq:Eulereq}.

Next, we consider Eq.~\eqref{eq:Eulereq} for small linearized perturbations about an equilibrium background configuration. We denote the background quantities by a subscript '$0$' and use a $\delta$ in front of the perturbed quantities, as in ${\bm u}={\bm u}_{0}+ \delta \bm u$. Let us recapitulate the arguments in Ref.~\cite{Flanagan:2006sb} that lead to the finding that the magnetic part of the perturbation equation for the fluid at the leading 1PN order simply reduces to the \emph{Newtonian} one augmented by a 1PN driving force from $\bm \zeta$.
We assume that the fluid perturbation is only generated by an external gravitomagnetic field $\bm A_\text{ext}$, which is of 1PN order. This implies that all perturbed quantities must also be of 1PN order, e.g., $\delta \bm u = \Order(c^{-2})$.
The perturbed magnetic Euler equation at 1PN order is then given by a perturbation of the Newtonian terms and $\bm \zeta$, that is, it derives solely from the terms shown in Eq.~\eqref{eq:Eulereq}.
The matter inside the perturbed neutron star is not static, and thus there is a gravitomagnetic field emanating from the neutron star. However, the perturbations sourcing this field are of 1PN order and the gravitomagnetic field equation is 1PN order, so the gravitomagnetic ``response'' field of the neutron-star perturbation is of 2PN order, which we can ignore for our purposes here. On the other hand, the fluid perturbations source a 1PN gravitoelectric field $\delta \phi$. Altogether, the 1PN harmonic-gauge perturbed field equations are given by
\begin{equation}\label{eq:PhiEOM}
\Delta \delta \bm A = 0, \qquad \Delta \delta \phi = 4\pi G \, \delta\rho .
\end{equation}
This implies that the perturbed gravitomagnetic potential only has contributions from sources external to the star $\delta \bm A = \bm A_\text{ext}$.
As mentioned above, we assume that there is no external gravitoelectric potential, $\phi_\text{ext} = 0$; we further comment on the 1PN gravitoelectric driving of magnetic modes in Sec.~\ref{sec:powercount}, finding that it is subleading at quadrupolar level.

Having identified the relevant field contributions, and terms at 1PN order, we now consider the perturbed fluid and no longer exhibit powers of $c$ explicitly. The background fluid is at rest in the corotating frame, ${\bm u}_{0}=0$. The external gravitomagnetic perturbation induces a Lagrangian fluid displacement ${\bm \xi}({\bm x},t)$ in the star, such that $\delta \bm u = \dot {\bm \xi}$. Calculating the perturbations to the Euler equation~\eqref{eq:Eulereq} and keeping only 1PN terms up to linear order in the perturbations and in the angular velocity leads to 
\begin{equation}
\label{eq:perturbedEuler}
\ddot{\bm \xi}+2 \bm \Omega \times {\dot{\bm \xi}}=-\frac{\bm \nabla \delta p}{\rho_{0}}+\frac{\bm \nabla p_{0}}{\rho_{0}^{2}} \delta \rho-\bm \nabla \delta\phi+\bm a_\text{ext} .
\end{equation}
The fluid acceleration induced by the external field $\bm a_\text{ext} = \delta{\bm \zeta}$ is given by
\begin{equation}
\label{eq:a_ext}
\begin{split}
  \bm a_\text{ext} =& -\dot{\bm A}_\text{ext}+ \bm \nabla [ ({\bm \Omega} \times {\bm x}) \cdot \bm A_\text{ext} ] .
\end{split}
\end{equation}
We emphasize that the above results are in the corotating frame; the analogous fluid perturbation equation in the inertial frame can be found in Eq.~(5.16) of Ref.~\cite{Flanagan:2006sb}.

Note that the background quantities in Eq.~\eqref{eq:perturbedEuler} refer to the zeroth order in the double expansion of post-Newtonian and rotational corrections: it is the equilibrium configuration computed for a Newtonian, nonrotating star. The feature that the background is identical to a nonrotating star at linear order in angular velocity is due to the fact that the Coriolis force on the background vanishes and the centripetal force is quadratic in $\bm \Omega$.
Hence the background quantities $\rho_0$, $p_0$ are spherically symmetric in our approximations.
In the following section, we obtain the Lagrangian for the perturbed Euler equation~\eqref{eq:perturbedEuler}.

Before proceeding, we highlight the following. To define the tidal deformability or Love number one studies the response of the star to the external field $\bm A_\text{ext}$ that is encoded in the induced gravitomagnetic field $\vct{A}$.
In our approximation, this response is sourced by the fluid perturbation $\vct{\xi}$ and is hence of 2PN order.
At that order, the nonlinear terms in the field equations lead to another source for the response field~\cite{Landry:2015cva,Poisson:2020mdi}.
Likewise, in our relativistic theory developed in Sec.~\ref{sec:relativistic}, we recover two contributions to the response, which can be attributed to the fluid displacement and field nonlinearities, respectively.
In the present section, however, we only consider the effect of the external field on the 1PN fluid displacement and focus instead on understanding the matter variables.

\subsection{Magnetic tidal Lagrangian for slow rotation}

We assume that the fluid is characterized by a simple temperature- and composition-independent equation of state of the form $p=p(\rho)$. Then the pressure perturbation is $\delta p = \delta \rho \, d p / d \rho$ and the forcing terms on the right hand side of Eq.~\eqref{eq:perturbedEuler} can be written as
\begin{equation}
\label{eq:pofrhoidentity}
-\frac{\bm \nabla \delta p}{\rho_{0}}+\frac{\bm \nabla p_{0}}{\rho_{0}^{2}} \delta \rho- \bm \nabla \delta\phi= - \bm \nabla\left(c_s^2\frac{\delta \rho}{\rho_0}+\delta \phi\right),
\end{equation}
where $c_s^2 = d p /  d \rho$ is the speed of sound.
We also use that $\delta \rho=-\bm \nabla \cdot \left(\rho_{0} \bm{\xi}\right)$, which follows from perturbing the Newtonian continuity equation $\dot \rho = - \bm \nabla \cdot (\rho \bm u)$, and a solution for $\delta \phi$ from its field equation~\eqref{eq:PhiEOM}.
Inserting the relation \eqref{eq:pofrhoidentity} into the equations of motion~\eqref{eq:perturbedEuler} yields
\begin{equation}
\label{eq:simplifiedEuler}
\ddot{\bm \xi}+2 \bm \Omega \times {\dot{\bm \xi}}=-\mathcal{D}\bm{\xi}+\bm a_\text{ext},
\end{equation}
where the linear operator $\mathcal{D}$ is defined as
\begin{equation}
\label{eq:Ddef}
\mathcal{D} \boldsymbol{\xi}=-\bm \nabla\left\{\left[\frac{c_{s}^{2}}{\rho_{0}}+4 \pi G \Delta^{-1}\right] \bm \nabla \cdot\left(\rho_{0} \boldsymbol{\xi}\right)\right\}.
\end{equation}
Note that $\mathcal{D}$ is the differential operator describing perturbations of a nonrotating Newtonian star, and effects of rotation are included explicitly as the Coriolis term on the left hand side of Eq.~\eqref{eq:simplifiedEuler}. The operator $\mathcal{D}$ is Hermitian under the product $\left\langle\boldsymbol{\xi}, \boldsymbol{\xi}^{\prime}\right\rangle=\int d^{3} x \rho_{0} \, \boldsymbol{\xi}^{*} \cdot \boldsymbol{\xi}^{\prime}$ \cite{1964ApJ...139..664C}. Hence, its eigenvectors $\bm \xi_{n \ell m}$---the normal modes---form an orthonormal basis with
\begin{equation}
 \label{eq:xinormalization}
\left\langle\boldsymbol{\xi}_{n \ell m}, \boldsymbol{\xi}_{n^{\prime} \ell^{\prime} m^{\prime}}\right\rangle = \delta_{n n^{\prime}} \delta_{\ell \ell^{\prime}} \delta_{m m^{\prime}} .
\end{equation}
Their eigenvalues $\bar{\omega}_{n \ell}^{2}$ are real
\begin{equation}
\label{eq:Doperator}
\mathcal{D} \boldsymbol{\xi}_{n \ell m}=\bar{\omega}_{n \ell}^{2} \boldsymbol{\xi}_{n \ell m}.
\end{equation}

We can decompose a generic fluid displacement $\boldsymbol{\xi}$ into this basis as
\begin{equation}
\label{eq:modesum}
\boldsymbol{\xi}=\sum_{n \ell m} q_{n \ell m}(t) \boldsymbol{\xi}_{n \ell m}(\boldsymbol{x}), \quad q_{n \ell m}=\left\langle\boldsymbol{\xi}_{n \ell m}, \boldsymbol{\xi}\right\rangle, 
\end{equation}
with time-dependent amplitudes $q_{n \ell m}(t).$ The fact that the fluid displacement is real, $\boldsymbol{\xi}=\boldsymbol{\xi^{*}}$, implies that $q_{n \ell m}^{*}=(-1)^{m} q_{n \ell-m}$ due to the analogous relation for the spherical harmonics, which arise because in the corotating frame and to linear order in the rotation, the modes are the same as for a nonrotating star and can be decomposed into vector spherical harmonics. In general, three types of vector harmonics contribute to the modes, each with different parity: the parity-even electric and radial harmonics, and the parity-odd magnetic-type ones~\cite{Thorne:1980ru}. For the gravitomagnetic tidal dynamics considered here, only the magnetic-type ($\mathcal{B}$) contributions $\vct{\xi}^{\cal B} = \sum q^{\cal B}_{n \ell m} \boldsymbol{\xi}^{\cal B}_{n \ell m}$ are relevant, and we henceforth drop the other contributions. The magnetic modes $\boldsymbol{\xi}^{\cal B}_{n \ell m}$ have the general decomposition into a radial dependence $\xi_{n \ell}^{\mathcal{B}}(r)$ and the magnetic vector spherical harmonic $\vct{Y}^{\cal B}_{\ell m}(\theta, \varphi)$ depending on the polar and azimuthal angles $(\theta$, $\varphi)$, see Appendix \ref{sec:appendix},
\begin{equation}
\label{eq:ximagnetic}
\bm{\xi}_{n \ell m}^{\cal B} = \xi_{n \ell}^{\mathcal{B}}(r) \,\vct{Y}^{\cal B}_{\ell m}(\theta, \varphi).
\end{equation}
The normalization of the real radial mode function $\xi_{n \ell}^{\mathcal{B}}$ follows from Eqs.~\eqref{eq:xinormalization} and \eqref{eq:YBnormal},
\begin{equation}\label{eq:modenorm}
    \int dr \, \rho_0 r^2 \xi_{n \ell}^{\mathcal{B}}(r) \xi_{n' \ell}^{\mathcal{B}}(r) = \delta_{n n^{\prime}} ,
\end{equation}
and is accompanied by a completeness relation,
\begin{equation}
    \sum_n \xi_{n \ell}^{\mathcal{B}}(r) \xi_{n \ell}^{\mathcal{B}}(r') = \frac{\delta(r-r')}{\rho_0 r^2} . \label{eq:modecomplete}
\end{equation}
Note that at linear order in spin (and for the perfect fluid used here), the radial functions $\xi_{n \ell}^{\mathcal{B}}(r)$ are rather degenerate \cite{1981A&A....94..126P} and we can pick them to be any complete basis of functions labeled by $n$.

An important property of the magnetic modes that directly follows from the definition of the vector spherical harmonic~\eqref{eq:defYB} is that $\bm \nabla  \cdot (\rho_0 \bm \xi^{\cal B}) = 0$ and $
  \mathcal{D} \boldsymbol{\xi}^{\cal B}_{n \ell m}=0. $
Equation~\eqref{eq:Doperator} then tells us that the magnetic modes of the nonrotating star all have zero frequency, 
\begin{equation}
    \bar{\omega}^{\cal B}_{n \ell}=0 .
\end{equation}
As a result, the equations of motion \eqref{eq:simplifiedEuler} simplify to 
\begin{equation}
\label{eq:eommagnetic}
\ddot{\bm\xi}^{\cal B}+2 \bm \Omega \times {\dot{\bm \xi}^{\cal B}}=\bm a_\text{ext}.
\end{equation}
The Lagrangian for these equations of motion \eqref{eq:eommagnetic} is
\begin{equation}
\label{eq:LDT}
L_{\mathrm{DT}}^{\cal B}=\frac{1}{2}\langle\dot{\boldsymbol{\xi}}^{\cal B}, \dot{\boldsymbol{\xi}}^{\cal B}\rangle-\langle\boldsymbol{\xi}^{\cal B}, \boldsymbol{\Omega} \times \dot{\boldsymbol{\xi}}^{\cal B}\rangle+\langle \bm a_\text{ext}, \boldsymbol{\xi}^{\cal B}\rangle .
\end{equation}

Note that, if $\phi_\text{ext}$ was nonzero, the magnetic modes do not couple to the gravitoelectric potential at Newtonian order, $\langle \bm \nabla \phi_\text{ext}, \boldsymbol{\xi}^{\cal B}\rangle = 0$.
Indeed, using the definition of the inner product, the coefficient $\langle \bm \nabla \phi_\text{ext}, \boldsymbol{\xi}^{\cal B}\rangle$ is given by
\begin{equation}
\label{eq:elzero}
 \int d^{3} x \rho_{0} \, \bm \nabla \phi_\text{ext} \cdot \boldsymbol{\xi}^{\cal B} = % \int d^{3} x \,\bm \nabla \cdot (\phi_\text{ext} \rho_{0}\boldsymbol{\xi})
-  \int d^{3} x \, \phi_\text{ext} \bm \nabla \cdot( \rho_{0}\boldsymbol{\xi}^{\cal B}) = 0,
\end{equation}
where we dropped a surface term since $\bm \xi^{\cal B}=0$ on the surface of the star and used that $\bm \nabla  \cdot (\rho_0 \bm \xi^{\cal B}) = 0$.
We further discuss the coupling to the gravitoelectric potential beyond Newtonian order in Sec.~\ref{sec:powercount}.

In the next section, we discuss an interesting application of this Lagrangian that illustrates the utility of our formalism to calculate the gravitomagnetic mode frequency for the slowly rotating neutron star.

\subsection{Gravitomagnetic mode frequency}\label{sec:NewtonMode}

The operator $\mathcal{D}$ associated with the nonrotating star has eigenvalues $(\bar{\omega}_{n \ell}^{\cal B})^2=0$. However, the Coriolis term will give rise to nonzero eigenvalues which correspond to the gravitomagnetic mode frequency. The calculation of the mode frequency does not require the external gravitomagnetic force term; it is sufficient to consider the free oscillations described by the Lagrangian from Eq.~\eqref{eq:LDT} 
\begin{equation}
\label{eq:LDTfree}
L_{\mathrm{DT, free}}^{\cal B}=\frac{1}{2}\langle\dot{\boldsymbol{\xi}}^{\cal B}, \dot{\boldsymbol{\xi}}^{\cal B}\rangle-\langle\boldsymbol{\xi}^{\cal B}, \boldsymbol{\Omega} \times \dot{\boldsymbol{\xi}}^{\cal B}\rangle.
\end{equation}
Using the decomposition into mode amplitudes~\eqref{eq:modesum} and aligning the angular momentum as $\vct{\Omega} = (0,0,\Omega)$, the terms in the Lagrangian simplify as follows:
\begin{subequations}
\label{eq:freeLDTtermsamplitude}
\begin{align}
\frac{1}{2}\langle\dot{\boldsymbol{\xi}}^{\cal B}, \dot{\boldsymbol{\xi}}^{\cal B}\rangle =& \sum_{n \ell m}\frac{1}{2} \dot{q}^{\cal B *}_{n \ell m}(t)\dot{q}^{\cal B}_{n \ell m}(t), \\
-\langle\boldsymbol{\xi}^{\cal B}, \boldsymbol{\Omega} \times \dot{\boldsymbol{\xi}}^{\cal B}\rangle =& \sum_{n \ell m}q^{\cal B *}_{n \ell m}(t)\dot{q}^{\cal B}_{n \ell m}(t)\frac{i m\Omega}{\ell(\ell+1)}.
\end{align}
\end{subequations}
To arrive at these expressions, we made use of Eqs.~\eqref{eq:modenorm}, \eqref{eq:YBnormal}, \eqref{eq:YBrelation}, and \eqref{eq:Yrelation}.
We insert the results \eqref{eq:freeLDTtermsamplitude} into the free Lagrangian ~\eqref{eq:LDTfree} and obtain the free equations of motion for the mode amplitudes $q^{\cal B}_{n \ell m}(t)$
\begin{equation}
\label{eq:eomfreeBmodes}
\ddot{q}^{\cal B}_{n \ell m}= \frac{2 i m\Omega}{\ell(\ell+1)}\dot{q}^{\cal B}_{n \ell m}.
\end{equation}

To determine the mode frequency, we make the ansatz $q_{n \ell m}(t)=C e^{-i\omega^{\cal B}_{\ell m}t}$. We choose the minus sign in the exponent so that  the modes with positive $m$ have negative frequencies (see below), which matches the usual conventions for mode expansions in the presence of an external tidal driving $\sim e^{-im\phi}$.  Eq.~\eqref{eq:eomfreeBmodes} and find that it is a solution for frequencies $\omega^{\cal B}_{\ell m}$---the mode frequencies---given by
\begin{equation}
\label{eq:Bmodefreq}
\omega^{\cal B}_{\ell m}=-\frac{2 m \Omega}{\ell(\ell+1)} .
\end{equation}
Our calculation makes explicit that the effect of rotation is to shift the mode frequency away from its nonspinning value of $\bar\omega^{\cal B}_{n\ell}=0$ to the finite value given in Eq.~\eqref{eq:Bmodefreq}. We recall that these gravitomagnetic mode frequencies are expressed in the corotating frame. 

An important point to note is that the force-free equations \eqref{eq:eomfreeBmodes} have an additional solution with $\omega^{\cal B}_{\ell m} = 0$, which describes a constant mode amplitude.
This zero-mode is associated with the trivial displacements analyzed in Ref.~\cite{1978ApJ...221..937F} (see also Ref.~\cite{Flanagan:2006sb}). That is, constant displacements in the gravitomagnetic modes corresponding to the transformation
\begin{equation}\label{eq:trivial}
    q^{\cal B}_{n \ell m} \rightarrow q^{\cal B}_{n \ell m} + \text{const}_{n\ell m} ,
\end{equation}
leave the macroscopic properties of the fluid (density, pressure, velocity) unchanged and hence must be physically inconsequential.

\subsection{Lagrangian in the symmetric tracefree basis}

Next, we transform from the description in terms of $(\ell,m)$ modes to an equivalent one in terms of symmetric-tracefree tensors. This is advantageous for making the connection to the relativistic effective action in the next section.

We start by decomposing the Lagrangian~\eqref{eq:LDT} into the normal modes, using that
\begin{subequations}
\begin{gather}
\langle\dot{\boldsymbol{\xi}}^{\cal B}, \dot{\boldsymbol{\xi}}^{\cal B}\rangle = \sum_{n \ell m' m} \dot{q}^{\cal B*}_{n \ell m'} \dot{q}^{\cal B}_{n \ell m} N_\ell^2 \mathcal{Y}^{*\ell m'}_{S_1...S_\ell}\mathcal{Y}^{\ell m}_{S_1...S_\ell} , \\
\begin{split}
\langle\boldsymbol{\xi}^{\cal B}, \boldsymbol{\Omega} \times \dot{\boldsymbol{\xi}}^{\cal B}\rangle = & \sum_{n \ell m'm}\frac{1}{\ell+1}q^{\cal B *}_{n \ell m'} \dot{q}^{\cal B}_{n \ell m} \Omega^{AB} \\
&\times N_\ell^2 \mathcal{Y}^{*\ell m'}_{AS_1...S_{\ell-1}} \mathcal{Y}^{\ell m}_{BS_1...S_{\ell-1}},
\end{split}
\end{gather}
\end{subequations}
where $N_\ell=\sqrt{4\pi \ell! /(2\ell+1)!!}$, and we have used the identities in Eqs.~\eqref{eq:YBnormal} and \eqref{eq:YBrelation} together with the normalization in Eq.~\eqref{eq:modenorm}.
The symmetric-tracefree tensors $\mathcal{Y}_{s_{1} s_{2} \ldots s_{l}}^{l m}$ are explained in Appendix~\ref{sec:appendix}.

We next evaluate the coupling to the external field from $\langle \bm a_\text{ext}, \boldsymbol{\xi}^{\cal B}\rangle$ and Eq.~\eqref{eq:a_ext}.
For this purpose, it is useful to expand the external fields in a Taylor series in $\vct{x}$ around the center at $\vct{x} = \vct{0}$.
The first ($\ell=0$) and second ($\ell=1$) term in this expansion can be ignored, since they can not have a physical effect in a local free-falling frame due to the equivalence principle.
Hence from now on we specialize to the quadrupole case ($\ell=2$) giving the leading-order driving force, i.e., it contributes the largest effect in a binary.
Recalling that $\bm \nabla  \cdot (\rho_0 \bm \xi^{\cal B}) = 0$, we see that the second term in Eq.~\eqref{eq:a_ext} does not contribute, since one can partially integrate the gradient under the integral, as in Eq.~\eqref{eq:elzero}.
The only contribution then comes from the magnetic-spherical-harmonic ($\vct{Y}_{2m}^{{\cal B}}$) part of $\vct{A}^\text{ext}$, so we write its Taylor expansion in $\vct{x}$ as (different by a conventional sign from Ref.~\cite{Flanagan:2006sb})
\begin{align}
A^\text{ext}_{I} &= - \sum_m \sqrt{\frac{2}{3}} N_2 \mathcal{B}_m(t) r^2 Y_{2m}^{{\cal B}\, I}(\theta, \varphi) + \dots \\
&= -\frac{2}{3} \epsilon_{IJK} \mathcal{B}_{KL}(t) x^{J} x^{L} + \dots ,
\end{align}
where we used the definition of the magnetic vector spherical harmonic in the form of Eq.~\eqref{eq:defYB2} and we encoded the external quadrupolar gravitomagnetic field in a symmetric-tracefree tensor $\mathcal{B}_{KL}(t) = N_2 \sum_m \mathcal{Y}^{2 m}_{KL} \mathcal{B}_m(t)$.
Note that the vacuum field equations $\Delta \vct{A}^\text{ext} = 0$ are satisfied.
Now we straightforwardly obtain (for $\ell=2$)
\begin{equation}
\langle \bm a_\text{ext}, \boldsymbol{\xi}^{\cal B}\rangle = \sum_{nm} \sqrt{\frac{2}{3}} N_2^2 I_{n} q^{\cal B}_{n 2m} \mathcal{Y}^{2m}_{IJ} \dot{\mathcal{B}}_{IJ},
\end{equation}
using $B_m = N_2 \mathcal{Y}^{*2 m}_{IJ} \mathcal{B}_{IJ}$ and Eqs.~\eqref{eq:YBnormal}, \eqref{eq:Ycontract}, and where we defined the gravitomagnetic overlap integral as~\cite{Flanagan:2006sb}
\begin{equation}
I_n=\int dr \, \rho_0 r^{4} \xi_{n2}^{\cal B} .
\end{equation}

The quadrupolar mode amplitudes can be written in terms of symmetric-tracefree tensors as
\begin{equation}
Q_{{\cal B} n}^{I J}(t)= \sum_{m} \sqrt{\frac{2}{3}} 2 N_2^2 I_{n} \mathcal{Y}_{I J}^{2 m} q^{\cal B}_{n 2 m}(t) ,
\end{equation}
where the prefactor is chosen to simplify the coupling term with the external field.
With this definition, and after dropping total time derivatives, the Lagrangian becomes
\begin{align}
\label{eq:LDTSTFmode}
L_{\mathrm{DT}}^{\cal B}=&\sum_n \frac{3}{16 N_2^2 I_n^2} \left(\dot{Q}_{{\cal B}n}^{I J} \dot{Q}_{{\cal B}n}^{I J} - \frac{2}{3} \Omega^{JK}\dot{Q}_{{\cal B}n}^{I J}Q_{{\cal B}n}^{KI} \right) \nnl
-\frac{1}{2} \mathcal{B}_{IJ} \sum_n \dot{Q}_{{\cal B}n}^{I J} .
\end{align}

\subsection{Effective Lagrangian}

As we are mainly interested in the bulk interaction of the star with the gravitomagnetic field rather than the dynamics of individual modes $Q_{{\cal B}n}^{I J}$ with different radial profiles (but identical mode frequencies), it is convenient to define an effective internal degree of freedom as 
\begin{equation}
  Q_{\cal B}^{IJ} = \sum_n Q_{{\cal B}n}^{I J}.  
\end{equation} 
Its equation of motion follows from those for the individual modes $Q_{{\cal B}n}^{I J}$ and is given by
\begin{equation}
\ddot{Q}_{{\cal B}}^{I J} + \frac{2}{3} \Omega^{K(I}\dot{Q}_{{\cal B}}^{J) K} = \frac{4}{3} N_2^2 I_{\cal B}^2 \dot{\mathcal{B}}_{IJ} ,
\end{equation}
where 
\begin{equation}
\label{eq:IcalB}
I_{\cal B}^2 \equiv \sum_n I_n^2 = \int dr \, \rho_0 r^6 .
\end{equation}
In the last step we used the completeness relation of the modes~\eqref{eq:modecomplete} together with the definition of the $I_n$.
This new overlap integral $I_{\cal B}$ only depends on the background density $\rho_0$; compare also to Eq.~(6.13) in Ref.~\cite{Landry:2015cva}.
It is directly related to the post-Newtonian magnetic tidal deformabilities, c.f., Eq.~(6.20) in Ref.~\cite{Landry:2015cva}.
We elucidate the concrete connection to the Love numbers in the relativistic case in Sec.~\ref{sec:matching} below. 

We can write the Lagrangian for the effective gravitomagnetic degrees of freedom as
\begin{equation}
\label{eq:LDTSTF}
L_{\mathrm{DT}}^{{\cal B}}=C^{\cal B}_{\dot{Q}^2}\dot{Q}_{{\cal B}}^{I J} \dot{Q}_{{\cal B}}^{I J} + C^{\cal B}_{\Omega Q \dot{Q}} \Omega^{JK}\dot{Q}_{{\cal B}}^{I J}Q_{{\cal B}}^{KI}-\frac{1}{2} \mathcal{B}_{IJ} \dot{Q}_{{\cal B}}^{I J} .
\end{equation}
Note that this action is not identical to Eq.~\eqref{eq:LDTSTFmode} yet describes a physically equivalent interaction of the star with the gravitomagnetic field.
Here, we have kept the coefficients of the interaction terms as general constants, as will become important for the relativistic extension discussed in the next section. For a neutron star approximated to 1PN order they take the values ($N_2^2 = 8\pi / 15$)
\begin{equation}\label{eq:CNewton}
C_{\dot{Q}^2}^{\cal B} = \frac{3}{16 N_2^2 I_{\cal B}^2} = \frac{45}{128 \pi I_{\cal B}^2} , \qquad C_{\Omega Q \dot{Q}}^{\cal B} = - \frac{2}{3} C^{\cal B}_{\dot{Q}^2}.
\end{equation}
Furthermore, since $\Omega^{IJ} = \text{const}$ for an isolated star, the action \eqref{eq:LDTSTF} exhibits a shift symmetry
\begin{equation}\label{trivialsym}
    Q_{{\cal B}}^{I J} \rightarrow Q_{{\cal B}}^{I J} + \text{const}^{IJ} ,
\end{equation}
as expected from the symmetry of the fluid under trivial displacements~\eqref{eq:trivial} of the gravitomagnetic mode amplitudes~\cite{1978ApJ...221..937F}.

The Lagrangian in Eq.~\eqref{eq:LDTSTF} is expressed in the corotating frame. The transformation to the inertial frame is accomplished similarly as discussed for the metric in the beginning of Sec.~\ref{sec:Newtaction}, e.g., $Q_{\cal B}^{IJ} = R_I{}^i R_J{}^j Q_{\cal B }^{ij}$. This implies that the time derivatives transform as
\begin{equation}
\dot Q_{\cal B}^{IJ} = R_I{}^i R_J{}^j \underbrace{ \Big( \dot{Q}_{\cal B }^{ij} - 2 \Omega^{k(i} Q_{\cal B }^{j)k} \Big) }_{\displaystyle \equiv Q'^{ij}_{\cal B}} ,
\end{equation}
where $\Omega^{ij} = R_I{}^i \dot{R}^{Ij}$.
In order to maintain the shift symmetry~\eqref{trivialsym}, we express the inertial-frame action using the quantity $Q'^{ij}_{\cal B}$ which is invariant under the symmetry.
The action then becomes
\begin{equation}
L_{\mathrm{DT}}^{\cal B}=C^{\cal B}_{\dot{Q}^2} Q'^{ij}_{\cal B} Q'^{ij}_{\cal B} + C^{\cal B}_{\Omega Q \dot{Q}} \Omega^{jk}Q'^{ij}_{\cal B}Q_{{\cal B}}^{ki} - \frac{1}{2} \mathcal{B}_{ij} Q'^{ij}_{\cal B} .
\end{equation}
The insights about the relevant effective degrees of freedom and the role of the shift symmetry will be important inputs for constructing the relativistic effective action, as we discuss in the next section.

%%%%%%%%%%%%%%%%%%%%%%%%%%%%%%%%%%%%%%%
\section{Relativistic effective action}
\label{sec:relativistic}
%%%%%%%%%%%%%%%%%%%%%%%%%%%%%%%%%%%%%%%

In this section, we go beyond the 1PN corotating-frame Lagrangian for the magnetic modes by developing a fully relativistic action along a worldline.
We follow an effective-field-theory approach \cite{Goldberger:2004jt,Goldberger:2007hy,Foffa:2013qca,Rothstein:2014sra,Porto:2016pyg,Levi:2018nxp}, and construct an ansatz for such an action from symmetries, keeping interaction terms only up to a certain order (accuracy) in some power counting, e.g, the multipole counting.
The coefficients of the resulting terms in this action depend on the internal structure of the star and are fixed through a matching calculation that we will discuss in Sec.~\ref{sec:matching} below. 
Similar effective-field-theory treatments of dynamical tides and tidal absorption can be found in Refs.~\cite{Goldberger:2005cd, Steinhoff:2016rfi}, of spin can be found in Refs.~\cite{Porto:2005ac, Levi:2010zu, Levi:2015msa}, and of spin-tides can be found in Refs.~\cite{Porto:2007qi, Endlich:2015mke}.
Recently, effective-field-theory calculations of tidal effects in scattering events have also come into focus~\cite{Kalin:2020mvi,Kalin:2020lmz}, see also Ref.~\cite{Bini:2020flp}, and Refs.~\cite{Cheung:2020sdj,Haddad:2020que,Cheung:2020gbf,Bern:2020uwk} for analogous work based on massive quantum fields or scattering amplitudes.

We consider here a worldline $z^\mu(\tau)$ as a macroscopic, coarse-grained, effective description for a compact star.
We write the action as an integral of a Lagrangian $L$ over the proper time $\tau$ here,
\begin{equation}
\mathcal{S} = \int d \tau \, L .
\end{equation}
The simplest example of the Lagrangian is $L = m_0 = \text{const}$ which describes a point mass and neglects tides and spin.
The oscillation modes of the star are represented by dynamical variables that evolve along the worldline.
We start with an analysis of the symmetries of the problem, and demand that the building blocks of the action transform irreducibly under the symmetries.

Equipped with such a relativistic effective action, it is straightforward to calculate the orbital dynamics and gravitational radiation~\cite{Goldberger:2004jt,Goldberger:2007hy,Foffa:2013qca,Rothstein:2014sra,Porto:2016pyg,Levi:2018nxp}, see Ref.~\cite{Levi:2017kzq} for a publicly available code and Refs.~\cite{Bini:2012gu,Banihashemi:2018xfb,Henry:2019xhg,Henry:2020ski,Henry:2020pzq} for applications to tidal effects.
The development of a relativistic effective action for gravitomagnetic tides is hence a crucial step towards more realistic waveform models for neutron stars.

\subsection{Spherical symmetry and dynamical variables}

The most important symmetry assumption is the spherical symmetry of the nonrotating star in equilibrium.
This symmetry means that tidal degrees of freedom, or any other quantity from which we build the action, can
be arranged into 3-dimensional symmetric-tracefree tensors, which transform
irreducibly under rotations [the SO(3) group]. For generic spinning stars, the spherical
symmetry is broken and only an axial symmetry remains.
However, since we treat the spin perturbatively we can still base
our description on the rotation symmetry and the symmetric-tracefree tensor representation of
the nonrotating case, as in the previous section.
Following the effective action approach for electric tides~\cite{Steinhoff:2016rfi}, we
consider a dynamical variable $Q^{\mu\nu}(\tau)$ along the worldline representing quadrupolar ($\ell=2$) oscillation modes.
This quantity should be symmetric $Q^{[\mu\nu]} = 0$, tracefree $Q^\mu{}_\mu = 0$, and should have
physical components in the rest-frame only, $Q^{\mu\nu} U_\nu = 0$. The 
rest-frame is aligned with the tangent to the worldline, i.e., the
normalized 4-velocity $U^\mu = \dot z^\mu$ with $U_\mu U^\mu = -1$ and in this section, an overdot denotes a derivative with respect to proper time $\dot ~ = d / d \tau$.
We thus have a description of the mode in the coordinate frame with indices $\mu, \nu \dots$ running through 0, 1, 2, 3.

It is convenient to express the action
manifestly in terms of only the physical components of all quantities. To achieve this, we
introduce a corotating (body-fixed) frame $\Lcov_I{}^\mu$ with labels $I,J,\dots=1,2,3$ and
assume orthonormality, $\Lcov_I{}^\mu \Lcov_{J\mu} = \delta_{IJ}$. The temporal part $\Lcov_0{}^\mu$ is aligned
with the rest-frame,
\begin{equation}\label{covframe}
\Lcov_0{}^\mu = U^\mu , \quad \text{or } \Lcov_I{}^\mu U_\mu = 0 .
\end{equation}
Hence $\Lcov_I{}^\mu$ contains three independent angular degrees of freedom, as expected.
The three independent components of the angular velocity are 
\begin{equation}
\Omega_I = \frac{1}{2} \epsilon_{IJK} \Omega^{JK}
\end{equation}
 with
\begin{equation}
  \Omega^{JK} = - \Omega^{KJ} = \frac{D\Lcov^J{}_\mu}{d\tau} \Lcov^{K\mu} ,
\end{equation}
and $D$ is the covariant differential, see also Refs.~\cite{Goldberger:2009qd,Endlich:2015mke}.
This is a covariant generalization of the Newtonian angular velocity~\eqref{eq:OmegaNewton}.
Now, the independent components of the dynamical quadrupole are given by a
symmetric tracefree 3-tensor in the corotating frame
\begin{equation}
  Q^{IJ} = \Lcov^I{}_\mu \Lcov^J{}_\nu Q^{\mu\nu} , \quad
  Q^{[IJ]} = 0 = Q^I{}_I .
\end{equation}
Other internal or tidal degrees of freedom can likewise be expressed as symmetric-tracefree tensors in the corotating frame, such as an ($\ell=3$) octupole $Q^{IJK}$ and higher
multipoles $Q^{IJK\dots}$.
In this section we omit the label $n$ for enumerating different (families of) modes for simplicity.

\subsection{Coordinate invariance and external fields}
Another important symmetry of the action is general coordinate invariance, which requires that external fields coupling to the worldline must be tensors on spacetime.
However, in our setup, the external fields such as the curvature tensor entering the action must also be expressed in the corotating frame, followed by a decomposition into symmetric-tracefree parts.
The curvature tensor $R_{\mu\nu\alpha\beta}$ can be decomposed
into the Weyl tensor $C_{\mu\nu\alpha\beta}$ and the Ricci tensor $R_{\mu\nu}$. The latter can be
removed from the worldline action through redefinitions of the metric
\cite{Goldberger:2004jt, Goldberger:2005cd}, which essentially corresponds to using the
vacuum field equations $R_{\mu\nu} = 0$.
Furthermore, the Weyl tensor can be decomposed into symmetric-tracefree
electric $E_{IJ}$ and magnetic $B_{IJ}$ parts as
\begin{align}
  E_{IJ} =& C_{I0J0}
            = \Lcov_I{}^\mu \Lcov_0{}^\alpha \Lcov_J{}^\nu \Lcov_0{}^\beta C_{\mu\alpha\nu\beta} , \\
  B_{IJ} =& *\!C_{I0J0}
            =\Lcov_I{}^\mu \Lcov_0{}^\alpha \Lcov_J{}^\nu \Lcov_0{}^\beta *\!C_{\mu\alpha\nu\beta} ,
\end{align}
with the dual $*\!C_{\mu\nu\alpha\beta} = 1/2 \eta_{\mu\nu\rho\sigma} C^{\rho\sigma}{}_{\alpha\beta}$ and the volume form $\eta_{\mu\nu\rho\sigma}$.
Notice that $\epsilon_{IJK} = \eta_{0IJK} = \Lcov_0{}^\mu \Lcov_I{}^\nu \Lcov_J{}^\rho \Lcov_K{}^\sigma \eta_{\mu\nu\rho\sigma}$.
Derivatives of the Weyl curvature can be decomposed into the tensors (for $s \geq 2$)
\begin{align}
  E_{K_1\dots K_s} =&
            \Lcov_0{}^\alpha \Lcov_0{}^\beta \Lcov_{(K_1}{}^{\mu_1} \dots \Lcov_{K_s)}{}^{\mu_s} \nabla_{\mu_3 \dots \mu_s} C_{\mu_1 \alpha \mu_2 \beta} , \\
  B_{K_1 \dots K_s} =&
            \Lcov_0{}^\alpha \Lcov_0{}^\beta \Lcov_{(K_1}{}^{\mu_1} \dots \Lcov_{K_s)}{}^{\mu_s} \nabla_{\mu_3 \dots \mu_s} *\!C_{\mu_1 \alpha \mu_2 \beta} .
\end{align}
Other components can be written as time derivatives via
$\Lambda_0{}^\mu \nabla_\mu \equiv D/d\tau$, see Ref.~\cite{Levi:2015msa} for more details.
We indicate the parity of the modes by a subscript: $Q_{\cal E}^{IJK\dots}$ are even parity (electric) and  $Q_{\cal B}^{IJK\dots}$ are odd
parity (magnetic) modes.

\subsection{Final set of building blocks for the action}
Based on the considerations above, the tensors entering the effective action are
\begin{equation}\label{variables}
  \epsilon_{IJK} , \quad
  \Omega^{I}, \quad
  Q_{\cal E}^{IJ\dots}, \quad
  Q_{\cal B}^{IJ\dots}, \quad
  E_{IJ\dots}, \quad
 B_{IJ\dots},
\end{equation}
together with their $\tau$-derivatives,  and the tensors $\delta_{IJ}$, $\delta^{IJ}$. Recall that $U^I = \Lambda^{I\mu} U_\mu  = 0$, which implies that the four-velocity cannot appear explicitly in the action.
All of these building blocks for the action are conveniently written in the corotating frame.

\subsection{Symmetry restrictions on the possible couplings}
Overall, we require the following symmetries from the effective worldline action, see also Ref.~\cite{Levi:2015msa}:
\begin{enumerate}
\item First, we require general coordinate invariance, required by general relativity.
\item Second, we require SO(3) symmetry of internal degrees of freedom due to the spherical symmetry of the
  body in the nonrotating limit. We already discussed one of the implications of this symmetry in the context of the building blocks of the action above.
\item Third, we require an ``external'' SO(3) symmetry of the corotating frame describing the orientation of the body.\footnote{A point-particle is characterized by an irreducible representation of the Poincar\'e group. The ``external'' SO(3) symmetry is the so called little group of that representation and is associated to the spin, see also \cite{Endlich:2015mke}.} In the case considered here, this SO(3) symmetry appears together with the internal SO(3) symmetry of 2. as one, which is very economic.
\item Fourth, we require spacetime parity invariance. In principle, weak interactions could violate this symmetry, however, they play a subdominant role for the structure of a neutron star. 
\item Fifth, we require time reversal invariance. This would be broken by dissipative effects such as tidal heating or fluid viscosity, which we neglect here.\footnote{An action-based treatment of dissipative tidal effects requires a more general approach, see Refs.~\cite{Goldberger:2009qd,Porto:2007qi,Goldberger:2012kf,Endlich:2015mke,Endlich:2016jgc,Galley:2009px,Galley:2012hx}. In this paper, we consider only conservative effects at the body scale.}
\item Sixth, we require shift symmetry of the gravitomagnetic dynamical tidal variables
\begin{equation}\label{shiftsym}
    Q_{{\cal B}}^{I J \dots} \rightarrow Q_{{\cal B}}^{I J \dots} + \text{const}^{IJ \dots} ,
\end{equation}
which is related to macroscopically unobservable fluid displacements~\cite{1978ApJ...221..937F}, as explained in the previous section around Eq.~\eqref{eq:trivial}.
This realization \eqref{shiftsym} of the symmetry was only demonstrated for $\Omega^{IJ} = \text{const}$ and might need to be amended in a more general setting.
\end{enumerate}

The consequences of the above symmetries on the terms in the action are the following.
The first three symmetries imply that interactions must be composed of
scalar contractions between the tensors in Eq.~(\ref{variables}) and their
$\tau$-derivatives. Symmetry 4 requires
interactions to contain an even number of odd-parity variables $\{ \epsilon_{IJK},
\Omega^I, Q^{IJ\dots}_{\cal B}, B_{IJ\dots}\}$. Symmetry 5 requires an even number of
variables that are odd under time reversal, which comprises $\Omega^I$ and $B_{IJ\dots}$, plus a number of extra $\tau$-derivatives.
Finally, symmetry 6 requires that terms in the action either depend on $Q_{{\cal B}}^{I J \dots}$ only via $\dot Q_{{\cal B}}^{I J \dots}$, or are of the form $Q_{{\cal B}}^{I J \dots}$ times a total $\tau$-derivative.
It is interesting to note that the terms that are not allowed by these symmetry requirements exactly match to what one would call selection rules, in analogy to atomic physics, for the overlap integrals appearing in Sec.~\ref{sec:Newtaction}.

We note that in contrast to Ref.~\cite{Levi:2015msa}, here, we assume time-reversal invariance and spell out parity invariance explicitly. Furthermore, we find it more convenient here to not include worldline reparametrization and spin-gauge invariance from the beginning. Those symmetries are important for calculating the post-Newtonian binary dynamics and can readily be introduced at a later stage by changing the evolution parameter from proper time $\tau$ to a generic affine parameter and  performing a boost of the corotating frame, as explained in  Sec.~3.2 of Ref.~\cite{Levi:2015msa}. 
For the purpose of post-Newtonian calculations, it is also convenient to promote the spin (conjugate to $\Omega^I$) to an additional dynamical variable via a Legendre transformation. The spin variable then absorbs the derivative coupling to the metric contained in  $\Omega^I$. This leads to considerable simplifications but will not be needed here.

\subsection{Power counting}\label{sec:powercount}

The next step in constructing an effective action is to include all interaction terms allowed by the symmetries listed above and up to a certain order in some power counting in a ratio of scales. Here, two distinct types of scale-ratios are relevant:  (i) for spatial scales we use the multipole counting in the ratio of the object's size and the radius of curvature of the external fields and (ii) for the time scales, we consider powers of the ratio of the various internal relaxation times and variations of the external tidal field. In the case where the external field is sourced by the companion in a binary system, (i) involves the orbital separation and (ii) involves multiples of the orbital period. These power countings must in general be treated as independent, e.g., for eccentric orbits.
We do not introduce an \emph{a priori} cutoff in a ratio of time scales, i.e., we allow for an arbitrary number of time derivatives.
However, we are interested here in the leading (quadrupolar) gravitomagnetic interaction, so we work to quadratic order in the odd-parity quadrupolar variables $Q_{{\cal B}}^{I J}$, $B_{IJ}$.

We note that a driving of magnetic modes by a quadrupolar gravitoelectric field is possible via  couplings such as $E_{IJ} \Omega_K \dot{Q}_{\cal B}^{IJK}$ and $E_{IJ} \Omega^I \dot{Q}_{\cal B}^J$, and similar couplings between electric and magnetic modes.
At 1PN order, these derive from a term of the form $\vct{\Omega} \times \vct{x} \, \dot \phi$ in the Euler equation~\eqref{eq:Eulereq}.
However, these couplings are suppressed by the angular velocity (flux dipole), making them effectively higher than quadrupolar order.
Likewise, the octupolar electric driving for the quadrupolar magnetic modes from the coupling $E_{IJK} \Omega_K \dot{Q}_{\cal B}^{IJ}$ is of higher order in the multipole counting.

Furthermore, we consider the angular velocity (spin) and its associated time (length) scale as an independent parameter and work to linear order in the angular velocity in the tidal interactions.
Finally, variable redefinitions can in general be used to remove some interaction terms from the action.
In particular higher-order time derivatives of the dynamical degrees of freedom on the worldline can be removed in this way \cite{Damour:1990jh}.
In the present case, this means that we can disregard $\dddot Q_{{\cal B}}^{I J}$ or $\dot \Omega^{IJ}$ or even higher time derivatives from the ansatz.

\subsection{Relativistic action in the corotating frame}

We now have all the inputs for deriving the relativistic tidal Lagrangian following the procedure outlined above.
The complete Lagrangian consists of a nontidal and a magnetic tidal part,
\begin{equation}
  L = L_\text{NT} + L_\text{DT}^{\cal B} .
\end{equation}
The nontidal part contains terms such as
\begin{equation}
  L_\text{NT} = - m_0 + C_{\Omega^2} \Omega^I \Omega^I + C_{E\Omega^2} E_{IJ} \Omega^I \Omega^J + \dots ,
\end{equation}
where $C_{\Omega^2}$ is related to the moment of inertia and $ C_{E\Omega^2}$ is related to the spin-induced quadrupole moment of the star~\cite{1967ApJ...150.1005H,Poisson:1997ha,Laarakkers:1997hb}.

Following the symmetries listed above and assuming a single type of quadrupolar gravitomagnetic modes $Q_{{\cal B}}^{I J}$, we obtain the following relativistic effective action in the corotating frame for dynamical magnetic tides to quadratic order in the tidal variables and linear order in the angular velocity of the star,
\begin{equation}
\begin{split}\label{LmagFull}
L_{\mathrm{DT}}^{{\cal B}} \approx C^{\cal B}_{\dot{Q}^2}\dot{Q}_{{\cal B}}^{I J} \dot{Q}_{{\cal B}}^{I J} + C^{\cal B}_{\Omega Q \dot{Q}} \Omega^{JK} \dot{Q}_{{\cal B}}^{I J} Q_{{\cal B}}^{KI}
-\frac{1}{2} B_{IJ} \dot{Q}_{{\cal B}}^{I J} \\
+ B_{IJ} \sum_{k=0}^\infty \bigg[ C_{BB^{(2k)}} \partial_\tau^{2k} \! B_{IJ}
+ C_{BB^{(2k+1)}\Omega} \Omega^{KI} \partial_\tau^{2k+1} \! B_{JK} \bigg] \! ,
\end{split}
\end{equation}
where $\partial_\tau = d / d \tau$. 
We choose the convention for the normalization of $Q^{IJ}_{{\cal B}}$ such that the coefficient in front of the third term is $-\frac{1}{2}$, as in the 1PN case.
Higher $\tau$-derivatives on $\Omega^{IJ}$ and $Q_{{\cal B}}^{I J}$ can be removed by variable redefinitions in the action.
The second term here is invariant under the shift symmetry \eqref{shiftsym} since $\dot{\Omega}^{IJ} \approx 0 + \Order(E, B)$.\footnote{Strictly speaking, the shift symmetry has to hold without using equations of motion. However, since one can shift the dynamical variables to remove time derivatives like $\dot{\Omega}$, one way to fix this is to amend the transformation rule~\eqref{shiftsym}. We leave such a more rigorous treatment for the case $\dot{\Omega} \neq 0$ for future work.}
We note that $B_{IJ} \approx A^\text{ext}_{K,L(I} \epsilon_{J)LK} / 2 \approx \mathcal{B}_{IJ}$ to 1PN order for a star at rest.
This can be checked by calculating the curvature tensor from the inertial-frame metric~\eqref{eq:metricinertial} and using its transformation property under coordinate changes to arrive at the corotating frame via Eq.~\eqref{eq:coordchange}.

The equations of motion for the mode amplitudes read
\begin{equation}\label{eq:relEOM}
    C^{\cal B}_{\dot{Q}^2}\ddot{Q}_{{\cal B}}^{I J} - C^{\cal B}_{\Omega Q \dot{Q}} \Omega^{K(I} \dot{Q}_{{\cal B}}^{J)K} = \frac{1}{4} \dot B_{IJ} .
\end{equation}
The source of the gravitational field equations, the energy momentum tensor, follows from a variation with respect to the metric.
This source can be decomposed into mass and flux multipoles, with the order corresponding to the number of spatial derivatives acting on the metric in the action.
Indeed, the star's current or flux quadrupole $\mathcal{J}^{IJ}$ in the corotating frame can be identified directly from the action~\cite{Banihashemi:2018xfb}: it is the quantity that couples to $B_{IJ}$ as $2/3 B_{IJ} \mathcal{J}^{IJ}$, and is given by
\begin{align}\label{fluxquadfull}
    \mathcal{J}^{IJ} \equiv \frac{3}{2} \frac{\delta L}{\delta B_{IJ}} &\approx - \frac{3}{4} \dot Q{}_{\cal B}^{IJ}
    %+ 3 C_{B^2} B_{IJ}
    + 3 \sum_{k=0}^\infty \bigg[ C_{BB^{(2k)}} \partial_\tau^{2k} \! B_{IJ} \nnl
    + C_{BB^{(2k+1)}\Omega} \Omega_{K(I} \partial_\tau^{2k+1} \! B_{J)K} \bigg] .
\end{align}
The flux quadrupole is the source term for the gravitomagnetic response field of the star.
The first contribution in Eq.~\eqref{fluxquadfull} is due to the odd-parity fluid perturbation described by $Q_{\cal B}^{IJ}$. In the absence of any additional fluid modes besides $Q_{\cal B}^{IJ}$, as we assume here, the remaining terms in Eq.~\eqref{fluxquadfull} may be interpreted as a nonlinear field contribution.
This interpretation is in agreement with the post-Newtonian analysis in Ref.~\cite{Landry:2015cva} and in the last section.

In the post-Newtonian approximation, e.g., for a bound binary system, each time derivative on $B_{IJ}$ leads to a further suppression of the term.
At leading order, we may therefore neglect all but the term without $\tau$-derivatives in the second line of Eq.~\eqref{LmagFull}.
By contrast, terms with time derivatives on $Q_{{\cal B}}^{I J}$ must be kept, since the fluid modes can be resonantly excited.
This yields the simplified action
\begin{equation}
\begin{split}\label{Lmag}
L_{\mathrm{DT}}^{{\cal B}} &\approx C^{\cal B}_{\dot{Q}^2}\dot{Q}_{{\cal B}}^{I J} \dot{Q}_{{\cal B}}^{I J} + C^{\cal B}_{\Omega Q \dot{Q}} \Omega^{JK}\dot{Q}_{{\cal B}}^{I J}Q_{{\cal B}}^{KI} \nl
-\frac{1}{2} B_{IJ} \dot{Q}_{{\cal B}}^{I J}
+ C_{B^2} B_{IJ} B_{IJ} ,
\end{split}
\end{equation}
with $C_{B^2} \equiv C_{BB^{(0)}}$.
%and flux quadrupole moment reads
%\begin{equation}
%    \mathcal{J}^{IJ} \approx - \frac{3}{4} \dot Q{}_{\cal B}^{IJ} + 3 C_{B^2} B_{IJ} . \label{fluxquad}
%\end{equation}
The equations of motion~\eqref{eq:relEOM} are not affected by this additional approximation.

\subsection{Relativistic action in the coordinate frame}

It is straightforward to rewrite the action in the coordinate frame using the transformation matrices $\Lambda_I{}^\mu$,
\begin{equation}
\begin{split}\label{LmagCoord}
L_{\mathrm{DT}}^{\cal B} &\approx C^{\cal B}_{\dot{Q}^2} Q'^{\mu\nu}_{\cal B} Q'^{{\cal B}}_{\mu\nu}
+ C^{\cal B}_{\Omega Q \dot{Q}} \Omega_{\nu\rho}Q'^{\mu\nu}_{{\cal B}} Q_{{\cal B} \mu}^{\rho} \nl
-\frac{1}{2} B_{\mu\nu} Q'^{\mu\nu}_{{\cal B}} + C_{B^2} B_{\mu\nu} B^{\mu\nu} ,
\end{split}
\end{equation}
where $\Omega^{\mu\nu} = \Lambda_I{}^\mu D \Lambda^{I\nu} / d \tau$. We have defined the quantity
\begin{equation}
\label{Qprimecoord}
Q'^{\mu\nu}_{\cal B} = \frac{D Q_{\cal B }^{\mu\nu}}{d \tau} + 2 \Omega^{(\mu}{}_\rho Q_{\cal B }^{\nu)\rho} , \end{equation}
which is invariant under the shift symmetry \eqref{shiftsym}. We note that even though we work to $\Order(\Omega)$, the dependence on the angular velocity in Eq.~\eqref{Qprimecoord} should not be expanded out, to ensure that the zero-mode is preserved. 
Also, since $\Lambda_I{}^\mu U_\mu = 0$, the variables in the coordinate-frame action are subject to constraints (or supplementary conditions),
\begin{equation}
    \Omega^{\mu\nu} U_\nu = 0, \qquad Q_{\cal B}^{\mu\nu} U_\nu = 0 .
\end{equation}
From this effective action~\eqref{LmagCoord}, one can follow Refs.~\cite{Levi:2015msa,Steinhoff:2016rfi,Levi:2017kzq} to work out the post-Newtonian description of a binary system by performing a Legendre transform in $\Omega^{\mu\nu}$ and $\dot{Q}_{{\cal B}}^{\mu\nu}$, introducing worldline reparametrization- and spin-gauge invariance, implementing a gauge fixing, deriving the Feynman rules, and calculating observables.
It is important that the post-Newtonian approximation is applied only at the scale of the binary.
That is, the neutron-star interior is treated in full general relativity and its internal structure is encoded in the coefficients of the effective action.

The post-Newtonian predictions for observables  depend on the coefficients in the action~\eqref{LmagCoord}.
The next important step is thus to match them for a given fully relativistic neutron-star model.
Before doing so, it is illustrative to compare Eq.~\eqref{LmagCoord} to the corresponding Lagrangian for dynamical, electric fundamental (f-)modes for a nonrotating star given by~\cite{Steinhoff:2016rfi} 
\begin{align}\label{eq:Lel}
L_\text{DT}^{\cal E} &\approx \frac{1}{4 \lambda \omega_f^2} \left[ \frac{D Q_{\cal E}^{\mu\nu}}{d \tau} \frac{D Q^\mathcal{E}_{\mu\nu}}{d \tau}
  - \omega_f^2 Q_{\cal E}^{\mu\nu} Q^\mathcal{E}_{\mu\nu} \right] - \frac{1}{2} E_{\mu\nu} Q_{\cal E}^{\mu\nu} ,
\end{align}
where $\omega_f$ is the f-mode frequency and $\lambda$ the electric quadrupolar Love number (tidal deformability).
The differences to the magnetic action \eqref{LmagCoord} are due to the different parity and time-reversal properties of the magnetic variables and the shift symmetry~\eqref{shiftsym}.
In particular, the latter implies the absence of a $Q_{\cal B} ^2$-term, meaning that the magnetic modes have zero frequency in the nonrotating case.
As a consequence, the adiabatic limit of the magnetic action is not immediately obvious, in particular when the nonrotating limit is taken at the same time (see also Ref.~\cite{Poisson:2020mdi}).
In fact, as we show below, in the nonrotating adiabatic case the coefficients in the action are connected to both the irrotational and static versions of the Love numbers.
Another difference to the gravitomagnetic case is that possible $E_{\mu\nu} E^{\mu\nu}$-terms in Eq.~\eqref{eq:Lel} are approximately negligible for the description of dynamical f-modes~\cite{Chakrabarti:2013lua,Steinhoff:2016rfi}.

%%%%%%%%%%%%%%%%%%%%%%%%%%%%%%%%%%%%%%%
\section{Matching the coefficients}
\label{sec:matching}
%%%%%%%%%%%%%%%%%%%%%%%%%%%%%%%%%%%%%%%

The relativistic effective action~\eqref{Lmag} for dynamical gravitomagnetic tides has an immediate connection to gravitational-wave observables.
The action can be directly used in a post-Newtonian approximate calculation of the binary dynamics as in Refs.~\cite{Bini:2012gu,Banihashemi:2018xfb,Henry:2019xhg,Henry:2020ski,Henry:2020pzq} or included in the effective one body model~\cite{Steinhoff:2016rfi,Hinderer:2016eia} to predict the effect of these tidal interactions on the gravitational waves from a binary inspiral.
This prediction generally depends on the coefficients in the effective action.
Hence the constants $C_{\dots}$ can be measured or constrained with gravitational-wave observations.

In order to link measurements to the nuclear physics of neutron stars, it is essential to theoretically calculate the coefficients in the action for relativistic neutron-star models.
In this section, we use matching arguments to relate the constants $C_{\dots}$ in Eq.~\eqref{Lmag} to the quadrupolar relativistic magnetic mode frequencies $\omega^{\cal B}_{2m}$ and the tidal deformabilities (Love numbers) $\sigma_\text{irr}$ and $\sigma_\text{stat}$.
Two distinct magnetic Love numbers have been defined in the literature for nonrotating neutron stars: negative irrotational ones $\sigma_\text{irr} < 0$~\cite{Damour:2009vw} and positive static ones~$\sigma_\text{static} > 0$ \cite{Binnington:2009bb}, which differ by the boundary conditions imposed in the fluid.
These can be computed numerically from linear perturbations of neutron stars, for $\omega^{\cal B}_{2m}$ see Refs.~\cite{Andersson:2000mf,Idrisy:2014qca,Kantor:2020dex,Lee:2002fp,Kokkotas:2015gea} and for $\sigma_{\dots}$ see Refs.~\cite{Damour:2009vw, Binnington:2009bb, Landry:2015cva, Landry:2015snx, Pani:2018inf}.

Note that the matching of tidal coefficients is inherently difficult and the definition of Love numbers may even be considered ambiguous \cite{Gralla:2017djj}. However, those ambiguities are expected to be comparable to rather small effects at 6PN order for magnetic tides (5PN order for electric tides). For the black-hole case, it is crucial to understand these subtleties since tidal effects, if nonzero, would be very small. Here, for neutron stars described by a perfect fluid equation of state, we can take a more heuristic approach to the matching as explained below. But when accounting for more realistic physics such as different classes of magnetic modes, more terms in the effective action become relevant and it is important to work out a more general and rigorous approach to the matching. This can be accomplished by matching the tidal parameters defined by coefficients in the effective action based on observables such as the binding energy, redshift, or scattering angle. We leave this for future work.

Finally, we note that quasi-universal relations, i.e., relations that are approximately independent of the nuclear equation of state, were studied for the magnetic Love numbers for neutron stars in Ref.~\cite{Delsate:2015wia}. The result was that these relations hold only to approximately $5\%$ ($10\%$) for irrotational (static) Love numbers; for irrotational Love numbers this was also found in recent follow-up work in Refs.~\cite{Gagnon-Bischoff:2017tnz, Pani:2015nua, Jimenez-Forteza:2018buh}. This should be compared to the nearly sub-percent-level universality for the electric-type Love numbers~\cite{Yagi:2013bca}, and opens interesting prospects for learning new information about the equation of state.

\subsection{Matching the static Love number}

We start by considering the case of no rotation $\Omega^I = 0$ and a static fluid $\dot Q_{\cal B}^{IJ} = 0$, and identify which coefficients can be fixed. 
The condition $\dot Q_{\cal B}^{IJ} = 0$ corresponds to a vanishing fluid velocity perturbation, which is the boundary condition leading to the static Love number $\sigma_\text{stat}$ \cite{Landry:2015cva}.
Furthermore, the Love numbers are defined in the adiabatic limit, meaning that one also neglects time derivatives of $B_{IJ}$.
In this case, the flux quadrupole \eqref{fluxquadfull} and the action \eqref{LmagFull} reduce to
\begin{equation}
\mathcal{J}^{IJ}_\text{stat} \approx 3 C_{B^2} B_{IJ} , \qquad
L_\text{stat}^{\cal B} \approx %C_{\dot{Q}^2} \dot Q_{\cal B}^{IJ} \dot Q^{\cal B}_{IJ}
  % + C_{\Omega Q \dot{Q}} \Omega_{IJ} Q_{\cal B}^I{}_K \dot{Q}_{\cal B}^{KJ}
  %-\frac{1}{2} B_{IJ} \dot Q_{\cal B}^{IJ}
  C_{B^2} B_{IJ} B_{IJ} .
  %+ C_{B\Omega Q} B_{IJ} \Omega^I{}_K Q_{\cal B}^{KJ} .
\end{equation}
In general, the magnetic quadrupolar Love numbers $\sigma \equiv \sigma_2$ can be defined either as the proportionality constant between the magnetic tidal field $B_{IJ}$ and the flux quadrupole moment $S^{IJ}$, or as a coefficient in the adiabatic tidal action,
\begin{equation}\label{eq:defLove}
  \mathcal{J}^{IJ} = 2 \sigma B^{IJ} , \quad \text{or} \quad
  L_\text{ad}^{\cal B} = \frac{2 \sigma}{3} B_{\mu\nu} B^{\mu\nu} .
\end{equation}
This definition holds for both the irrotational $\sigma_\text{irr}$ and static $\sigma_\text{stat}$ Love numbers.
Since we consider a static fluid in this subsection, we identify $\sigma$ with the static Love number $\sigma_\text{stat}$ here.
Comparing either definition in Eq.~\eqref{eq:defLove} with the above relations, we can match $C_{B^2}$ as
\begin{equation}
    C_{B^2} = \frac{2}{3} \sigma_\text{stat} .
\end{equation}

\subsection{Matching the irrotational Love number}

The irrotational Love number $\sigma_\text{irr}$ for a nonrotating fluid $\Omega^I = 0$ can be obtained by keeping the time/frequency dependence general, here in particular of $Q_{\cal B}^{IJ}$, and taking the adiabatic limit after arriving at a master equation for the perturbations~\cite{Damour:2009vw, Pani:2018inf}.

The equation of motion~\eqref{eq:relEOM} for $Q_{\cal B}^{IJ}$ simply reads $\ddot Q_{\cal B}^{IJ} = \dot B_{IJ} / (4 C^\mathcal{B}_{\dot{Q}^2})$ in the nonrotating case.
Integrating this equation with respect to $\tau$ and dropping the integration constant, which would lead to a permanent flux quadrupole, we obtain
\begin{equation}
  \dot Q_{\cal B}^{IJ} \approx \frac{B_{IJ}}{4 C_{\dot{Q}^2}^\mathcal{B}} .
\end{equation}
Inserting this into Eq.~\eqref{fluxquadfull} or Eq.~\eqref{LmagFull}, taking the adiabatic limit by dropping time derivatives of $B_{IJ}$, and comparing to the definition of the Love number \eqref{eq:defLove} leads to
\begin{equation}
    \sigma_\text{irr} = \frac{3 C_{B^2}}{2} - \frac{3}{32 C_{\dot{Q}^2}^\mathcal{B}} .
\end{equation}

To make the matching of the coefficients in the action more transparent, we split the Love number into its matter $\sigma_M$ and field $\sigma_F$ contributions (analogous to the post-Newtonian case in Ref.~\cite{Poisson:2020mdi}),
\begin{equation}
\label{eq:sigmaMFdef}
    \sigma_M \equiv \sigma_\text{irr} - \sigma_\text{stat} < 0, \qquad \sigma_F \equiv \sigma_\text{stat} > 0 .
\end{equation}
These quantities are related to coefficients in the nonrotating action by
\begin{equation}
    C_{\dot{Q}^2}^\mathcal{B} = - \frac{3}{32 \sigma_M} > 0, \qquad C_{B^2} = \frac{2 \sigma_F}{3} > 0 .
\end{equation}
Indeed, we must have $C^\mathcal{B}_{\dot{Q}^2} > 0$ for consistency since the energy of the modes must be bounded from below.
To the leading post-Newtonian order, the static and irrotational Love numbers are approximately related by $\sigma_\text{stat} \approx - 3 \sigma_\text{irr}$~\cite{Landry:2015cva}, which implies that $\sigma_F \approx - 3 \sigma_M / 4$.
However, both Love numbers can be obtained numerically from full relativistic perturbations of neutron stars and there is no immediate need to resort to these approximations.
Note that calculations based on the post-Newtonian theory for the mode amplitudes can be upgraded to a fully relativistic treatment of the interior by replacing the coefficients in the action~\eqref{eq:CNewton} by the relativistic values just identified.
Comparing to Eq.~\eqref{eq:CNewton}, we hence find for the overlap integral $I_{\cal B}$ the relativistic expression
\begin{equation}
    I_{\cal B}^2 = - \frac{15}{4 \pi} \sigma_M = \frac{15}{4 \pi} ( \sigma_\text{stat} - \sigma_\text{irr} ) .
\end{equation}
Similarly, for the mode frequencies one should use relativistic results, which we discuss now.

\subsection{Matching the mode frequency}

To identify the mode frequencies we follow the method discussed in Sec.~\ref{sec:NewtonMode} .
For this purpose, we first express the corotating-frame equations of motion~\eqref{eq:relEOM} of the mode amplitudes for an isolated neutron star at rest in the spherical-harmonic basis, using $Q_{\cal B}^{IJ} = N_2 \sum_m \mathcal{Y}_{IJ}^{2m} Q^{\cal B}_m$, where $N_2 = \sqrt{8 \pi / 15}$, and similarly for $B_{IJ}$.
Assuming that the (constant) angular velocity is aligned with the z-axis $\vct{\Omega} = (0, 0, \Omega)$, the equations of motion~\eqref{eq:relEOM} read
\begin{equation}\label{eq:relEOMlm}
2 C^{\cal B}_{\dot{Q}^2} \ddot Q^{\cal B}_m + i m \Omega C^{\cal B}_{\Omega Q \dot{Q}} \dot Q^{\cal B}_m = \frac{1}{2} \dot B_m ,
\end{equation}
which follows from contracting the equations of motion with $2 N_2 \mathcal{Y}_{IJ}^{*2m}$ and using~\eqref{eq:Ycontract}, \eqref{eq:Yrelation}.
The reality conditions $Q_{\cal B}^{IJ} = Q_{\cal B}^{*IJ}$ imply that $Q_m^{\cal B *} = (-1)^m Q^{\cal B}_{-m}$ and similarly for $B_m$.
We can determine the corotating-frame mode frequencies $\omega^{\cal B}_{2m}$ by making the ansatz $Q_m = C e^{-i\omega^{\cal B}_{\ell m}\tau}$ in the free ($B_m = 0$) equations of motion, which leads to
\begin{equation}
\label{eq:omegahatdef}
    \hat{\omega}_{\cal B} \equiv \frac{\omega^{\cal B}_{2m}}{m \Omega} = \frac{C^{\cal B}_{\Omega Q \dot{Q}}}{2 C^{\cal B}_{\dot{Q}^2}} < 0 ,
\end{equation}
and to the zero-mode $\omega^{\cal B}_{2m} = 0$.
We can thus match the coefficient $C^{\cal B}_{\Omega Q \dot{Q}}$ as
\begin{equation}
    C^{\cal B}_{\Omega Q \dot{Q}} = 2 C^{\cal B}_{\dot{Q}^2} \hat{\omega}_{\cal B} = - \frac{3 \hat{\omega}_{\cal B}}{16 \sigma_M} < 0 .
\end{equation}
This completes the matching of all coefficients in the simplified effective Lagrangian~\eqref{Lmag}.
We discuss in the next section how the matching of all coefficients in the Lagrangian in Eq.~\eqref{LmagFull} could be achieved based on the tidal response function, as a foundation for future work on the matching in more complicated scenarios. 
%Note that when matching to relativistic (damped) quasi-normal mode frequencies, then the damping might need to be taken into account~\cite{Maggiore:2007nq}.

The relativistic frequencies $\omega^{\cal B}_{2m}$ were computed in Ref.~\cite{Andersson:2000mf} for polytropes and in Ref.~\cite{Idrisy:2014qca}  
for nuclear physics-based equations of state.
In the slow-rotation limit and for typical neutron-star compactnesses $G M/(Rc^2)\lesssim 0.2$, they may differ by up to 15\% from the Newtonian estimate $\hat{\omega}_{\cal B} \approx - 1 / 3$, c.f., Eq.~\eqref{eq:Bmodefreq}. The relativistic corrections introduce a dependence of the mode frequency on the equation of state. 
See Ref.~\cite{Kokkotas:2015gea} for a review of the magnitude and equation-of-state dependence of relativistic and higher-order rotational corrections.
For superfluid stars, a second family of r-modes emerges, which has interesting consequences~\cite{Kantor:2020dex, Lee:2002fp}.
Such a scenario can be described in our effective theory framework by introducing effective dynamical tidal variables $Q_{{\cal B} n}^{IJ}$ and corresponding coefficients for each family of modes, where $n$ labels the family of modes.
We briefly elaborate on the matching in this scenario based on the response function below.

%%%%%%%%%%%%%%%%%%%%%%%%%%%%%%%%%%%%%%%
\section{Tidal response and Love operator}
\label{sec:response}
%%%%%%%%%%%%%%%%%%%%%%%%%%%%%%%%%%%%%%%

The quadrupolar tidal response function, or quadrupole propagator on the worldline, was introduced for the description of dissipative tides in the effective-field-theory context in Refs.~\cite{Goldberger:2005cd,Porto:2007qi}.
For conservative electric dynamical tides, it was further explored numerically in Refs.~\cite{Chakrabarti:2013lua,Chakrabarti:2013xza}.
An extension of this work to magnetic tides and slow rotation would be valuable.

In this section, we derive the frequency-domain linear tidal response function from the relativistic effective action~\eqref{LmagFull}.
We also discuss the response in three different limiting regimes: when the external frequency is smaller than the rotation frequency (pre-resonance), when they are comparable (near-resonance), and when the external frequency is larger than the rotation frequency (post-resonance).
In the time domain, the response is encoded in a tensorial linear integral operator, which may be dubbed the Love operator.

\subsection{The response function}

It is most transparent to study the tidal response in spherical-harmonic basis and frequency domain.
We hence further transform to the Fourier domain by using
\begin{equation}
  Q^{\cal B}_m (\tau) = \int \frac{d \omega}{2 \pi} \tilde Q^{\cal B}_m(\omega) e^{- i \omega \tau} ,
\end{equation}
and similarly for other tensors like $B_m$.
Based on the transformed flux quadrupole $\tilde{\mathcal{J}}_m$, we define the linear tidal response function $\tilde{F}^{\cal B}_m(\omega)$ of the neutron star in the corotating frame as a generalization of the Love number~\eqref{eq:defLove},
\begin{equation}
\tilde{\mathcal{J}}_m = 2 \tilde F^{\cal B}_m \tilde B_m .
\end{equation}
This response is the relativistic analog of the Love tensor from Ref.~\cite{Poisson:2020ify}, see also Ref.~\cite{LeTiec:2020spy} for the black-hole case.
The flux quadrupole in the spherical-harmonic decomposition in the frequency domain $\tilde{\mathcal{J}}_m$ can be obtained from Eq.~\eqref{fluxquadfull},
\begin{multline}
    \tilde{\mathcal{J}}_m = \frac{3}{4} i \omega \tilde Q^{\cal B}_m
      + 3 \tilde{B}_m \sum_{k=0}^\infty (i \omega)^{2k} \bigg[ C_{BB^{(2k)}} \\
  - \frac{m \omega \Omega}{2} C_{BB^{(2k+1)}\Omega} \bigg] .
\end{multline}
To obtain its explicit expression requires a solution for $\tilde Q^{\cal B}_m$, which we obtain by writing its equations of motion~\eqref{eq:relEOMlm} in the frequency domain,
\begin{equation}
\frac{3}{16 \sigma_M} \omega ( \omega - m \Omega \hat{\omega}_{\cal B} )\tilde Q^{\cal B}_m = - \frac{i \omega}{2} \tilde B_m .
\end{equation}
%\begin{equation}\label{Qsol}
%  \tilde Q_m = \frac{i \tilde B_m}{4 \omega C_{\dot{Q}^2} + 2 m \Omega C^{\cal B}_{\Omega Q \dot{Q}}} .
%\end{equation}
Solving for $\tilde Q^{\cal B}_m$ and using the above expression for $\tilde{\mathcal{J}}_m$, we arrive at the response function
\begin{align}
\label{eq:response}
    \tilde F^{\cal B}_m &= \sigma_M \frac{\omega}{\omega - m \Omega \hat{\omega}_{\cal B}} \nl
  + \frac{3}{2} \sum_{k=0}^\infty (i \omega)^{2k} \bigg( C_{BB^{(2k)}}
  - \frac{m \omega \Omega}{2} C_{BB^{(2k+1)}\Omega} \bigg) \nonumber \\
&\approx \sigma_M \frac{\omega}{\omega - m \Omega \hat{\omega}_{\cal B}} + \sigma_F ,
\end{align}
where the last line refers to the simplified action~\eqref{Lmag}.
With this, the action can be written as
\begin{align}\label{SDT}
S_\text{DT}^{\cal B} &\approx \int \frac{d\omega}{2\pi} \sum_{m = -2}^2 \frac{2}{3} \tilde F^{\cal B}_m \tilde B_m \tilde B_m^* .
\end{align}
Note that although we are working only to linear order in the spin, we do not expand the denominator in the response ~\eqref{eq:response}. The reason is similar to the textbook example of an anharmonic oscillator~\cite{Landau1976Mechanics}, where one perturbatively expands all terms at the level of the equations of motion yet leaves any denominators of the solution unexpanded. This is crucial in order to preserve essential features of the dynamics, i.e., poles at resonances.
For the same reason, it is important to keep the shift symmetry \eqref{shiftsym} without expanding in spin, so that the zero-frequency mode is preserved.

The corotating-frame response function \eqref{eq:response} can be extended to the case of several mode families by summing over the contributions from several $Q_{{\cal B} n}^{IJ}$,
\begin{equation}
\begin{split}\label{eq:multiresponse}
    \tilde F^{\cal B}_m &= \sum_n \sigma_{Mn} \frac{\omega}{\omega- m \Omega \hat{\omega}_{{\cal B}n}} \nl
  + \frac{3}{2} \sum_{n=0}^\infty (i \omega)^{2n} \bigg( C_{BB^{(2n)}}
  - \frac{m \omega \Omega}{2} C_{BB^{(2n+1)}\Omega} \bigg) ,
\end{split}
\end{equation}
where $32 \sigma_{Mn} C^{\cal B}_{n\dot{Q}^2} = -3$ and $16 \sigma_{Mn} C^{\cal B}_{n\Omega Q \dot{Q}} = -3 \hat{\omega}_{{\cal B}n}$.
In the presence of more than one family of modes, and in order to fix all coefficients in Eq.~\eqref{LmagFull}, a more general matching procedure than outlined above is necessary.
This could be accomplished through a numerical investigation of the magnetic tidal response $\tilde F^{\cal B}_m$ based on relativistic linear perturbation theory, similar to the nonrotating electric case in Ref.~\cite{Chakrabarti:2013lua}.
A fit of such a numerical result for $\tilde F^{\cal B}_m$ to Eq.~\eqref{eq:multiresponse} should in principle fix all the (linear, conservative) tidal coefficients:
The behavior of the response around its poles fixes the number of mode families and their coefficients $C^{\cal B}_{n\dot{Q}^2}$, $C^{\cal B}_{n\Omega Q \dot{Q}}$, while the global frequency dependence fixes the $C_{BB^{(2k)}}$, $C_{BB^{(2k+1)}\Omega}$.
Such a matching of the response could be further improved and made rigorous by basing it on gauge-invariant observables, as mentioned above.

\subsection{Time-domain response and Love operator}

Returning to the case of a single family of modes, we next show how to transform the response~\eqref{eq:response} back to the time-domain and bypass the problems encountered in Ref.~\cite{Poisson:2020mdi}. 
For this purpose, we work in the symmetric tracefree basis and choose retarded boundary conditions by setting $\omega \rightarrow \omega + i \epsilon$ with the limit $\epsilon \rightarrow 0^+$ implied.\footnote{We note that the denominator describing the propagator is linear in $\omega$ instead of quadratic, the latter being the more familiar case in field theory. This inhibits one to directly pick the Feynman prescription $\omega^2 \rightarrow \omega^2 + i \epsilon$ for the propagator/boundary conditions here. Advanced boundary conditions correspond to $\omega \rightarrow \omega - i \epsilon$.}
It is useful then to further shift the frequency by $\omega \rightarrow \omega + m \Omega \hat \omega_{\cal B}$, which introduces a phase $e^{-i m \Omega \hat \omega_{\cal B} \tau}$ from the Fourier transform.
Now, in the spherical harmonic basis, that phase can be absorbed by a shift of the azimuthal angle $\varphi$, i.e., a rotation around the spin axis.
After this chain of transformations, it is straightforward to perform the Fourier integral.
However, one can follow a more direct approach, also based on using rotations to conveniently simplify expressions, which we delineate in more detail below.

We start from the equations of motion~\eqref{eq:relEOM} for $Q_{\cal B}^{IJ}$ and write them in terms of the matter contribution to the flux quadrupole $\mathcal{J}_M^{IJ} \equiv - 3 \dot Q_{\cal B}^{IJ} / 4$ as
\begin{equation}\label{eq:fluxeom}
\frac{1}{4 \sigma_M} C^{\cal B}_{\dot{Q}^2} \left(\dot{\mathcal{J}}_M^{I J} - 2 \hat \omega_{\cal B} \Omega^{K(I}\mathcal{J}_M^{J) K} \right) = \frac{1}{2} \dot{B}_{IJ} .
\end{equation}
 This can be simplified by performing a rotation around the spin axis with
\begin{equation}
    \mathcal{J}_M^{IJ} = \bar R^I{}_K \bar R^J{}_L \bar{\mathcal{J}}_M^{KL},
    \quad \bar R = \exp(- \hat \omega_{\cal B} *\!\vct{\Omega} \tau) .
\end{equation}
We note the identity $\bar R(\tau) = \exp(- \hat \omega_{\cal B} *\vct{\Omega}^\mathsf{T} \tau)^\mathsf{T} = \bar R^\mathsf{T} (- \tau)$ since $*\vct{\Omega}{}^\mathsf{T} = - *\vct{\Omega}$, where ${}^\mathsf{T}$ denotes the matrix transpose.
Hence $\bar R$ is indeed a rotation matrix, $\bar R^{-1} = \bar R^\mathsf{T}$.
This rotation turns Eq.~\eqref{eq:fluxeom} into the simpler relation
\begin{equation}
 \bar R^I{}_K \bar R^J{}_L \dot{\bar{\mathcal{J}}}_M^{KL} = 2 \sigma_M \dot{B}_{IJ} ,
\end{equation}
where we recall our assumption that $\vct{\Omega} = \text{const}$.
The retarded solution for $\mathcal{J}_M^{IJ}$ is then given by
\begin{equation}
\label{timedomainJ}
\begin{split}
 \mathcal{J}_M^{IJ}(\tau) &= 2 \sigma_M \bar R^I{}_K(\tau) \bar R^J{}_L(\tau) \nl
 \times\int_{-\infty}^\tau d\tau' \, \bar R^A{}_K(\tau') \bar R^B{}_L(\tau') \frac{d B_{AB}(\tau')}{d\tau'} .
\end{split}
\end{equation}

With this solution at hand, assuming that $B_{AB}(-\infty) = 0$, and noting that $\bar R(\tau) \bar R^\mathsf{T}(\tau') = \bar R(\tau-\tau')$, we can define a tensorial magnetic Love operator $\hat \sigma^{IJKL}_\text{ret}$ such that
\begin{equation}
    \mathcal{J}^{IJ} = 2 \hat \sigma^{IJKL}_\text{ret} B_{KL}
    \equiv 2 \int_{-\infty}^\infty d\tau' \, F^{IJKL}_{{\cal B},\text{ret}}(\tau - \tau')  B_{KL}(\tau') ,
\end{equation}
with the time-domain response given by
\begin{equation}
\begin{split}\label{eq:simpleresponse}
F^{IJKL}_{{\cal B},\text{ret}}(\tau) &\approx \sigma_M \hat \delta^{IJAB} \frac{d\left( \Theta(\tau) \bar R^A{}_K(\tau) \bar R^B{}_L(\tau) \right)}{d \tau} \nl
+ \sigma_F \hat \delta^{IJKL} \delta(\tau) .
\end{split}
\end{equation}
Here, $\Theta(\tau)$ is the Heaviside step function which implements the retarded boundary conditions, and the symmetric-tracefree projector is given by
\begin{equation}
    \hat \delta^{IJKL} = \delta^{I(K} \delta^{L)J} - \frac{1}{3} \delta^{IJ} \delta^{KL}.
\end{equation}
This result \eqref{eq:simpleresponse} is specialized to the simplified action~\eqref{Lmag}. For the more general action in \eqref{LmagFull}, the retarded solution~\eqref{timedomainJ} continues to apply but with the response 
\begin{align}
F^{IJKL}_{{\cal B},\text{ret}}(\tau) &= \sigma_M \hat \delta^{IJAB} \frac{d\left( \Theta(\tau) \bar R^A{}_K(\tau) \bar R^B{}_L(\tau) \right)}{d \tau} \nnl
  + \frac{3}{2} \sum_{k=0}^\infty \bigg[ C_{BB^{(2k)}} \hat \delta^{IJKL} \partial_\tau^{2k} \delta(\tau) \nl
    + C_{BB^{(2k+1)}\Omega} \hat \delta^{IJA(K} \Omega^{L)A} \partial_\tau^{2k+1} \delta(\tau) \bigg]. \nonumber
\end{align}

Finally, we note that the action can be written as
\begin{align}
  S_\text{DT}^{\cal B} &= \int d\tau \, \frac{2}{3} B_{IJ} \hat{\sigma}^{IJKL} B_{KL} \\
  &= \int d\tau d\tau' \, \frac{2}{3}  B_{IJ}(\tau) F^{IJKL}_{\cal B}(\tau - \tau') B_{KL}(\tau') ,
\end{align}
which is nonlocal in time.
However, this action only encodes the time-symmetric part of the dynamics (the integrand can be symmetrized under $\tau \leftrightarrow \tau'$) and does not correspond to retarded boundary conditions~\cite{Galley:2012hx}.

\subsection{Asymptotic limits of the response function in the inertial frame}

We next discuss the frequency-domain response for the simplified effective Lagrangian in Eq.~\eqref{Lmag} involving a single mode family in the inertial frame.
The frequency in the inertial frame follows from that in the corotating frame via the relation 
\begin{equation}
\tilde \omega = \omega + m \Omega.
\end{equation}
Likewise, the inertial-frame mode frequency is $\tilde \omega^{\cal B}_{2m}  = \omega^{\cal B}_{2m}  + m \Omega$.
The gravitomagnetic response can then be expressed as
\begin{equation}
    \tilde F^{\cal B}_m \approx \sigma_M \frac{\tilde \omega - m \Omega}
  {\tilde \omega - \underbrace{(1+\hat{\omega}_{\cal B}) m \Omega}_{\displaystyle \tilde \omega^{\cal B}_{2m}} } + \sigma_F ,
\end{equation}
where $\hat{\omega}_{\cal B}$ is defined in \eqref{eq:omegahatdef} and $\sigma_{M,F}$ is defined in \eqref{eq:sigmaMFdef}.
We note that the adiabatic limit $\tilde \omega \rightarrow 0$ and the nonrotating limit $\Omega\rightarrow 0$ of the response do not commute~\cite{Poisson:2020mdi}.
Physically, this is not a problem since neither the neutron-star rotation frequency $\Omega$ nor the frequency of the external tidal field $\tilde \omega$ in a binary system is ever exactly zero.
What matters is the relation between $\tilde{\omega}$ and $\Omega$.
Away from the mode resonances, the response behaves as
\begin{equation}
    \tilde F^{\cal B}_m \approx \begin{cases}
    \displaystyle \frac{\sigma_M}{1+\hat{\omega}_{\cal B}} + \sigma_F & \text{for } |\tilde{\omega}| \ll |\Omega|, m \neq 0 , \\
    \sigma_M + \sigma_F = \sigma_\text{irr} & \text{for } |\tilde{\omega}| \gg |\Omega| \text{ or } m = 0 .
    \end{cases}
\end{equation}
However, we note that a proper treatment of the post-resonance regime $|\tilde{\omega}| \gg |\Omega|$ requires more care, in particular an analysis of the mode damping after resonant excitation.

%%%%%%%%%%%%%%%%%%%%%%%%%%%%%%%%%%%%%%%
\section{Summary and discussion}
\label{sec:summary}
%%%%%%%%%%%%%%%%%%%%%%%%%%%%%%%%%%%%%%%

Summarizing our findings, quadrupolar magnetic dynamical tides are approximately described by an effective action in the corotating frame given by
\begin{equation}
\begin{split}
L_{\mathrm{DT}}^{{\cal B}} &\approx -\frac{3}{32 \sigma_M} \left( \dot{Q}_{{\cal B}}^{I J} \dot{Q}_{{\cal B}}^{I J} + 2 \hat{\omega}_{\cal B} \Omega^{JK}\dot{Q}_{{\cal B}}^{I J}Q_{{\cal B}}^{KI} \right) \nl
-\frac{1}{2} B_{IJ} \dot{Q}_{{\cal B}}^{I J}
+ \frac{2 \sigma_F}{3} B_{IJ} B_{IJ} ,
\end{split}
\end{equation}
where
\begin{equation}
    \sigma_M \equiv \sigma_\text{irr} - \sigma_\text{stat}, \quad
    \sigma_F \equiv \sigma_\text{stat} , \quad
    \hat{\omega}_{\cal B} \equiv \frac{\omega^{\cal B}_{2m}}{m \Omega} ,
\end{equation}
$\sigma_\text{irr,stat}$ are the irrotational and static relativistic magnetic tidal deformabilities and $\omega^{\cal B}_{2m}$ are the relativistic mode frequencies.
These frequencies $\omega^{\cal B}_{2m}$ are linear in the magnetic spherical-harmonic number $m$ and angular velocity $\Omega$.
Hence $\hat{\omega}_{\cal B}$ is indeed independent of $m$ and $\Omega$.
We stress that for this action, the matching of the coefficients does not assume a post-Newtonian or low-compactness approximation of the neutron-star interior.
(In the low-compactness limit, $\hat{\omega}_{\cal B} \approx - 1 / 3$ and $\sigma_F \approx - 3 \sigma_M / 4$.)

However, the above model may become insufficient for a realistic inclusion of the microphysics of the neutron star.
For instance, the presence of superfluidity implies several families of magnetic modes \cite{Kantor:2020dex,Lee:2002fp,Yoshida:2003hc} and hence a more involved matching of the tidal parameters, as explained above.
Yet these complications also open new prospects for extracting precious information on neutron-star structure from gravitational waves.

The response function in the corotating frame is 
\begin{equation}
    \tilde F^{\cal B}_m \approx \sigma_M \frac{\omega }
  { \omega -  m \Omega\hat{\omega}_{\cal B} } + \sigma_F .
\end{equation}
In the inertial frame, where the frequency is $\omega+m\Omega$, the limiting forms of the response for $|\omega+m\Omega|/|\omega|\to 0,\infty$ (away from resonance) are 
\begin{equation}
    \tilde F^{\cal B}_m \approx \begin{cases}
    \displaystyle \frac{\sigma_\text{irr}+\hat{\omega}_{\cal B}\sigma_\text{stat}}{1+\hat{\omega}_{\cal B}},  &  |\tilde{\omega}| \ll |\Omega|, m \neq 0 , \\
    \sigma_\text{irr}, &  |\tilde{\omega}| \gg |\Omega| \text{ or } m = 0 .
    \end{cases}
\end{equation}
The main effect of a resonance is a phase shift in the gravitational waves, see the seminal work of Ref.~\cite{Flanagan:2006sb} for detailed calculations and Ref.~\cite{Ma:2020oni} for data-analysis implications.
This phase shift crucially depends on the overlap integral $I_{\cal B}$, which we are able to determine relativistically (for generic compactness) here,
\begin{equation}
    I_{\cal B}^2 = - \frac{15}{4 \pi} \sigma_M = \frac{15}{4 \pi} ( \sigma_\text{stat} - \sigma_\text{irr} ) .
\end{equation}
In addition, using a convenient set of rotations to simplify the equations leads to a tensorial magnetic Love operator characterizing the time-domain response as defined by the relation between flux quadrupole and gravitomagnetic field
\begin{equation}
    \mathcal{J}^{IJ} = 2 \hat \sigma^{IJKL}_\text{ret} B_{KL}
    \equiv 2 \int_{-\infty}^\infty d\tau' \, F^{IJKL}_{{\cal B},\text{ret}}(\tau - \tau')  B_{KL}(\tau') ,
\end{equation}
with the response given by
\begin{equation}
\begin{split}
F^{IJKL}_{{\cal B},\text{ret}}(\tau) &\approx \sigma_M \hat \delta^{IJAB} \frac{d\left( \Theta(\tau) \bar R^A{}_K(\tau) \bar R^B{}_L(\tau) \right)}{d \tau} \nl
+ \sigma_F \hat \delta^{IJKL} \delta(\tau) .
\end{split}
\end{equation}

It is interesting to consider the importance of the effects derived here for waveforms. The impact of the dynamical gravitomagnetic effects on the gravitational-wave phasing can be approximated as a sudden jump in the gravitational-wave phase $\psi$ at the resonance as~\cite{Flanagan:2006sb}
\begin{equation}
\label{eq:phasejump}
    \psi^{\rm dyn.\, res.}=\Theta(f-f_{\rm res})\left(1-\frac{f}{f_{\rm res}}\right)\Delta\Phi_{\rm res},
\end{equation}
where $f$ is the gravitational-wave frequency. The coefficient $\Delta\Phi_{\rm res}$, which sets the size of the jump and depends on $I_{\cal B}^2$ and $\hat\omega_{\cal B}$ among other binary parameters, is given explicitly in Eqs.~(5.37) in Ref.~\cite{Flanagan:2006sb} (the derivation being based on a post-Newtonian star). The instantaneous effect of the resonance on the phase is small. However, it occurs early in the inspiral, at frequencies $f_{\rm res}\sim \tilde \omega^B_{2m} / (m \pi)$ proportional to the spin frequencies $\Omega$. This implies that resonances can occur over a wide range of gravitational-wave frequencies from below 10~Hz to a few hundred hertz. The information about these effects thus accumulates over numerous gravitational-wave cycles. As an order-of-magnitude estimate, the net change in the gravitational-wave phase scales as $\Delta \psi\sim 0.05 R^4 \Omega^{2/3} $ for equal masses~\cite{Flanagan:2006sb}. Measuring these signatures from gravitomagnetic mode resonances is thus an important scientific opportunity with  third-generation detectors such as the Einstein Telescope and Cosmic Explorer, and must be taken into account to avoid biases in the measured parameters~\cite{Ma:2020oni}. For loud signals, these modes could also have an impact for measurements with current detectors as they further improve in their sensitivity to binary inspirals.  

The general form of the resonance effects on gravitational waves~\eqref{eq:phasejump} applies for any kind of tidal resonance that occurs early during the inspiral, both gravitoelectric and gravitomagnetic. The kinds of tides are encoded in the coefficient $\Delta \Phi$. These dynamical resonance effects on the phase~\eqref{eq:phasejump} are approximated by a step function in frequency and are hence very different from smooth-in-$f$ post-Newtonian contributions. In general, matter effects in binary systems are described by distinct perturbative expansions from the post-Newtonian one (e.g., the multipole expansion), with dimensionless parameters characterizing finite size effects as outlined in our discussion on the effective action and power counting in Sec.~\ref{sec:powercount}. For simplicity, assigning fiducial post-Newtonian orders to all physical effects is nevertheless often used to describe terms in the phasing with different powers of $f$, though formally this applies only to black holes. For instance, when considering tidal effects specialized to the adiabatic limit, the Fourier-domain phasing has the expansion
\begin{multline}
\label{eq:adiabaticphase}
    \psi^{\rm adiabatic}=\frac{3}{128\eta x^{5/2}}\bigg[1+a_{1{\rm PN}}x+O(x^{3/2}) \\
    \qquad -\frac{39}{2}\tilde \Lambda x^5+(\delta \Lambda+\tilde \Sigma)x^6+\ldots \bigg] ,
\end{multline}
where $x=[\pi f (M_1+M_2)]^{2/3}$ is a dimensionless frequency parameter, $\tilde \Lambda$ and $\delta \Lambda$ are dimensionless combinations of the individual gravitoelectric Love numbers $\lambda_{1,2}$ and the masses $M_{1,2}$ of the binary (characterizing adiabatic gravitoelectric tidal effects), and $\tilde \Sigma$ is a similar weighted average of the gravitomagnetic Love numbers $\sigma_{1,2}$. Effective post-Newtonian orders are attributed to each contribution according to the powers of $x$ involved, e.g., the leading-order gravitoelectric effects scale effectively as 5PN terms would, and magnetic effects start at effectively 6PN in this scheme.
The coefficients $\Lambda$ and $ \Sigma$ differ by nearly two orders of magnitude, i.e., the adiabatic magnetic effect is much smaller than even the subleading gravitoelectric effect.

However, the dynamical resonance effects induce nearly sudden changes at a particular frequency, as in Eq.~\eqref{eq:phasejump}, and do not fit even the fiducial post-Newtonian counting scheme of the adiabatic effects. They are a distinct phenomenon that can lead to a significantly larger imprint on the gravitational waves than suggested by the adiabatic limit. Thus, even though the contributions from \emph{adiabatic} gravitomagnetic effects~\eqref{eq:adiabaticphase} are very small, the \emph{dynamical} effects~\eqref{eq:phasejump} can be much more significant: Reference \cite{Ma:2020oni} demonstrates that including such dynamical effects in waveform models (and using quasi-universal relations) could improve constraints on certain tidal parameters by about two orders of magnitude for third-generation detectors. Furthermore, these gravitomagnetic resonances are not resolvable in numerical-relativity simulations of binary systems due to the length and timescales involved. This makes accurate \emph{analytical} modeling of these effects critically important. The analytical results depend on strong-field effects and the microphysics of the neutron stars in parametrized form through the overlap integral and the resonance frequency. This makes them broadly applicable to any type of compact object, and useful for tests of exotic objects, black holes, and gravity.

%%%%%%%%%%%%%%%%%%%%%%%%%%%%%%%%%%%%%%%
\section{Conclusions}
\label{sec:conclusions}
%%%%%%%%%%%%%%%%%%%%%%%%%%%%%%%%%%%%%%%

The observation of gravitational waves from binary neutron stars opens up exciting opportunities for exploring matter at supra-nuclear density in their interiors.
This requires understanding how the nuclear physics of neutron-star matter translates into tidal effects during the long inspiral phase of the binary, which constitutes a substantial part of the observed gravitational-wave signal.
In this paper, we made important progress on this topic by investigating how relativistic gravitomagnetic tides of neutron stars can be modeled with an effective action, both in the highly dynamical regime close to oscillation-mode resonances and away from resonance.

To gain intuition, we started from the 1PN description of a slowly rotating isolated neutron star, composed of an idealized fluid, in the presence of a gravitomagnetic tidal field. We derived a Lagrangian formulation of the linearized perturbations to the Euler equations for the fluid displacement. We used this Lagrangian to calculate the gravitomagnetic mode frequencies and to develop an effective action for composite degrees of freedom characterizing the gravitomagnetic interactions of the star. A crucial finding was a symmetry of the action under shifts of the dynamical mode degrees of freedom.

The major result of this paper is the fully relativistic effective action for gravitomagnetic tidal effects for slowly rotating neutron stars that we developed. We started from symmetry principles to construct the terms in the action, where the shift symmetry played an essential role. Each of these interaction terms comes with undetermined coefficients that encode the neutron-star structure. We demonstrated how the most important coefficients in the action match to the magnetic Love numbers and mode frequencies of the neutron star, showing that both kinds of magnetic Love numbers have physical relevance. We provide a relativistic expression for the overlap integral in terms of the Love numbers that may be used to improve estimates for the gravitational-wave phase shift when a neutron-star binary inspirals through magnetic-mode resonances. We also discussed several interesting dynamical consequences and unusual features compared to the gravitoelectric case, including the frequency-domain response function and time-domain Love operator.

An important goal for future work is to construct waveform models for gravitational waves from binary inspirals based on our effective action. The tidal coefficients in the action directly characterize the potentially measurable parameters in gravitational waves. These coefficients are related to the magnetic Love numbers and mode frequencies, which contain valuable information to better understand the extreme states of matter inside neutron stars.
Another target for future work is to formulate the matching in terms of the tidal response function, which would allow a generalization to more realistic microphysics, e.g., the inclusion of several mode families in the presence of a superfluid.

\acknowledgements

This work was supported by the Netherlands Organization for Scientific Research (NWO). T.H. acknowledges support from the NWO sectorplan, the DeltaITP, and NWO Projectruimte grant GW-EM NS.

\appendix

%%%%%%%%%%%%%%%%%%%%%%%%%%%%%%%%%%%%%%%
\section{Useful formulas}
\label{sec:appendix}
%%%%%%%%%%%%%%%%%%%%%%%%%%%%%%%%%%%%%%%

Let us collect here some useful formulas from symmetric-tracefree tensor formalism, see, e.g., Refs.~\cite{Thorne:1980ru,Damour:1990gj,Blanchet:1985sp}.
One can change from a symmetric-tracefree tensor basis (indices $s_1 \ldots s_\ell$) to a spherical-harmonic basis labeled by $(\ell,m)$ using the symbol $\mathcal{Y}_{s_{1} \dots s_{\ell}}^{\ell m}$, as in
\begin{equation}
Y_{\ell m}(\theta, \phi)=\mathcal{Y}_{s_{1} s_{2} \ldots s_{\ell}}^{\ell m} n^{s_{1}} n^{s_{2}} \ldots n^{s_{\ell}} ,
\end{equation}
where $n^{i} = x^{i} / r$ is the unit radial vector.
Here, the ordinary spherical harmonics $Y^{\ell m}(\theta, \varphi)$ depend on the polar and azimuthal angles $(\theta$, $\varphi)$.
The following holds,
\begin{equation}\label{eq:Ycontract}
    N_\ell^2 \mathcal{Y}_{s_{1} \ldots s_{\ell}}^{*\ell m'} \mathcal{Y}_{s_{1} \ldots s_{\ell}}^{\ell m} = \delta_{m'm} ,
\end{equation}
where $N_\ell=\sqrt{4\pi \ell! /(2\ell+1)!!}$.
Furthermore, Eq.~(2.26) of Ref.~\cite{Thorne:1980ru} leads to
\begin{equation}\label{eq:Yrelation}
    e_z^i \epsilon_{ijk} N_\ell^2 \mathcal{Y}_{j s_{1} \ldots s_{\ell-1}}^{*\ell m'} \mathcal{Y}_{k s_{1} \ldots s_{\ell-1}}^{\ell m} = \frac{i m}{\ell} \delta_{m'm} ,
\end{equation}
where $\vct{e}_z = (0,0,1)$.

A very useful integral formula is Eq.~(2.3) in Ref.~\cite{Thorne:1980ru},
\begin{equation}
\int d \Omega \, n_{i_{1}} \ldots n_{i_{2 \ell}}=\frac{N_\ell^2}{\ell!} ( \delta_{i_{1} i_{2}} \delta_{i_{3} i_{4}} \ldots \delta_{i_{2 \ell-1} i_{2 \ell}}+\ldots ) ,
\end{equation}
where the sum runs over all combinations of indices, and the integral is zero for an odd number of $\vct{n}$-vectors in the integrand.
For instance,
\begin{equation}
    \int d \Omega \, n^i n^j n^k n^\ell = \frac{4\pi}{15} ( \delta_{ij} \delta_{k\ell} + \delta_{ik} \delta_{j\ell} + \delta_{i\ell} \delta_{jk} ) .
\end{equation}
From the general integral formula, one can derive an extension of Eqs.~(2.5), (2.6) of Ref.~\cite{Thorne:1980ru}:
for any two symmetric-tracefree tensors $A_{s_1 \dots s_\ell}$, $B_{s_1 \dots s_{\ell'}}$ with $\ell \geq \ell'$ the following holds,
\begin{subequations}\label{eq:n2STF}
\begin{align}
    &\int d\Omega \, n^i n^j A_{s_1 \ldots s_\ell} n^{s_1}\ldots n^{s_\ell} B_{r_1 \ldots r_{\ell'}} n^{r_1}\ldots n^{r_{\ell'}} \nonumber \\
    &\quad= \frac{N_{\ell+1}^2}{(\ell+1)}
    \big[ \delta_{ij} A_{s_1 \ldots s_\ell} B_{s_{1}\ldots s_{\ell}} \nonumber \\
    &\qquad + 2 \ell A_{s_1\ldots s_{\ell-1}(i} B_{j)s_{1}\ldots s_{\ell-1}} \big] \quad \text{if $\ell = \ell'$}  ,  \\
    &\quad= N_\ell^2 A_{ij s_1 \ldots s_{\ell'}} B_{s_{1}\ldots s_{\ell'}} \quad \text{if $\ell = \ell' + 2$} , \\
    &\quad= 0 \quad \text{else}.
\end{align}
\end{subequations}
The symmetric-tracefree property of $A$, $B$ is crucial here, which means any two same indices (a trace) give zero, e.g., $A_{\ldots i \dots i \ldots}=0$.

The magnetic vector spherical harmonics $\vct{Y}^{\ell m}_{\cal B}$ play an important role in the present paper.
They are defined as
\begin{equation}\label{eq:defYB}
\vct{Y}_{\ell m}^{\cal B}(\theta, \varphi) = \frac{1}{\sqrt{\ell(\ell+1)}} \bm{x} \times \bm\nabla Y^{\ell m} ,
\end{equation}
or more explicitly in components as
\begin{equation}\label{eq:defYB2}
Y_{\ell m}^{{\cal B} i}(\theta, \varphi) = \frac{\sqrt{\ell}}{\sqrt{\ell+1}} \epsilon^{ijk} n^j \mathcal{Y}_{k s_{1} \ldots s_{\ell-1}}^{\ell m} n^{s_1} \ldots n^{s_{\ell-1}}  .
\end{equation}
The prefactor is chosen to satisfy the normalization
\begin{equation}\label{eq:YBnormal}
    \int d \Omega \, \vct{Y}_{\ell' m'}^{{\cal B} *} \cdot \vct{Y}_{\ell m}^{\cal B} = \delta_{\ell'\ell} \delta_{m'm} .
\end{equation}
This can be shown with the help of Eq.~\eqref{eq:Ycontract}, recalling that the $\mathcal{Y}_{s_{1} \dots s_{\ell}}^{\ell m}$ are symmetric tracefree in the indices $s_i$, and the relation
\begin{align}\label{eq:YBrelation}
    &\int d \Omega \, Y_{\ell' m'}^{{\cal B} * i} Y_{\ell m}^{{\cal B} j} =
    \frac{N_{\ell}^2 \delta_{\ell'\ell}}{\ell+1} \mathcal{Y}_{a s_{1} \ldots s_{\ell-1}}^{*\ell m'} \mathcal{Y}_{b s_{1} \ldots s_{\ell-1}}^{\ell m} \nonumber \\ 
    &\quad \times [ \ell \delta_{ij} \delta_{ab} - (\ell-1) \delta_{ia} \delta_{jb} - \ell \delta_{ib} \delta_{ja} ] ,
\end{align}
which in turn follows from Eq.~\eqref{eq:n2STF}, where one shifts $\ell \rightarrow \ell-1$ and identifies $A_{s_1 \dots s_{\ell-1}}^{k,lm} = \mathcal{Y}_{k s_{1} \ldots s_{\ell-1}}^{\ell m}$, and similar for $B_{s_1 \dots s_{\ell'-1}}^{k,l'm'}$.

\bibliography{main,main_auto}

\end{document}